\documentclass[12 pt]{article}
\usepackage[pdftex]{graphicx}
\usepackage{epstopdf}
\usepackage{amsmath}
\usepackage{epic}
\usepackage{eepic}
\usepackage{subfigure}
\usepackage{enumitem}

\newcommand{\debug}[1]{\mbox{\tt #1}}
\renewcommand{\debug}[1]{}              
\newcommand{\cmd}[1]{}

\newcommand{\daddcontentsline}[3]{\addcontentsline{#1}{#2}{#3}}

\newcommand{\nc}{\newcommand}
\nc{\bk}{\backslash}

\nc{\vb}{\verb}
\nc{\fix}[1]{\fbox{\sc #1}}

\newcommand{\iB}{\begin{itemize}}
\newcommand{\iE}{\end{itemize}}
\newcommand{\eB}{\begin{enumerate}}
\newcommand{\eE}{\end{enumerate}}
\newcommand{\dB}{\begin{description}%
}
\newcommand{\dE}{\end{description}}
\newcommand{\hB}{\begin{alphlist}}
\newcommand{\hE}{\end{alphlist}}
\newcommand{\hBa}{\begin{alphlista}}
\newcommand{\hEa}{\end{alphlista}}
\newcommand{\dBa}{\begin{description}%
\itemsep = -4.0pt}
\newcommand{\dBb}{\begin{description}%
\itemsep = -4.0pt \itemindent = -.5cm}
\newcommand{\2}{\>\>}
\newcommand{\3}{\>\>\>}

\newcommand{\bif}{\mbox{\boldmath $if \;\; $}}

\newcommand{\belse}{\mbox{\boldmath $else \;\; $}}

\newcommand{\bcase}{\mbox{\boldmath $case \;\; $}}

\newcommand{\TB}[1]{%
	\renewcommand{\tname}{\sname #1}%
	\daddcontentsline{lot}{table}{\debug{Theorem%
	\thetheorem \fbox{\sname #1}}}\begin{theorem}\slabelx{#1}%
	\cmd{TB} \  }
\newcommand{\TE}{\end{theorem}%
	\cmd{TE} }
\newcommand{\LB}[1]{%
	\renewcommand{\tname}{\sname #1}%
	\daddcontentsline{lot}{table}{\debug{Lemma \thelemma
	\fbox{\sname #1}}}\begin{lemma}\slabelx{#1}%
	\cmd{LB} \  }
\newcommand{\LE}{\end{lemma} %
	\cmd{ LE } }
\newcommand{\DefB}[1]{%
	\renewcommand{\tname}{\sname #1}%
    \daddcontentsline{lot}{table}{\debug{Definition%
    \thedefinition \fbox{\sname #1}}}\begin{definition}\slabelx{#1}%
	\cmd{DefB} \ }
\newcommand{\DefE}{\end{definition} \cmd{DefE} }

\newcommand{\PB}{{\em Proof:\/\ }\cmd{ PB} }

\newcommand{\EB}{\begin{equation}\cmd{EB}}
\newcommand{\EE}[1]{ \debug{\fbox{\sname #1}}%
	\label{\sname #1} \end{equation}%
	\cmd{EE} }

\newcommand{\NotB}[1]{ \begin{notation}\debug{\fbox{\sname #1}}%
	\label{\sname #1} %
	\cmd{NotB} }
\newcommand{\NotE}{\end{notation}\cmd{NotE}}

\newcommand{\EAB}{\begin{eqnarray}\cmd{EAB}}
\newcommand{\EAE}[1]{ \debug{\fbox{\sname #1}}%
    \label{\sname #1} \end{eqnarray}%
	\cmd{EAE} }
\nc{\EABs}{\begin{eqnarray*}\cmd{EABs}}
\nc{\EAEs}{\end{eqnarray*}\cmd{EAEs}}
\newcommand{\prB}[1]{\begin{problem}\slabelx{#1}\cmd{prB}%
	\daddcontentsline{lot}{table}{\debug{Problem
	\theproblem \fbox{\sname #1}}}\rm}
\newcommand{\prE}{\end{problem}\cmd{prE}  }

\newcommand{\FIGSli}[4]{\FB%
\vbox to #4in%
{\special{figure #1}}%
\end{figure}}
\newcommand{\sname}{}
\newcommand{\pname}{}
\newcommand{\tname}{}

\newcounter{ctr}
\renewcommand{\thectr}{\alph{ctr}}

\newenvironment{alphlist}{%
\begin{list}{\thectr)}{\usecounter{ctr}%
\topsep=0pt}%
}%
{\end{list}}
\newenvironment{alphlista}{%
\begin{list}{\thectr)}{\usecounter{ctr}%
\itemsep = -4.0pt \topsep=0pt}%
}%
{\end{list}}

\newcommand{\dlabel}[1]{\debug{\fbox{\tiny #1}}\cmd{dlabel}\label{#1}}
\newcommand{\dcite}[1]{\cite{#1}\debug{[#1]}\cmd{dcite}}
\newcommand{\dref}[1]{\ref{#1}\debug{[#1]}\cmd{dref}}

\newcommand{\href}[1]{phase#1()\cmd{href}}

\newcommand{\dlabelx}[1]{\debug{\fbox{\tiny #1}}\label{#1}}

\newcommand{\slabelx}[1]{\debug{\fbox{\tiny \sname #1}}%
	\label{\sname #1}}
\newcommand{\plabelx}[1]{\debug{\fbox{\tiny \pname #1}}%
	\label{\pname #1}}

\newcommand{\textwidtha}{6.25in}

\newcommand{\ls}[1]%
{%
\large%
\normalsize}

\newcounter{protblock}
\newcounter{line}[protblock]
\renewcommand{\theline}{\mbox{\Alph{protblock}\arabic{line}}}
\newcommand{\nl}[1]{\\ \refstepcounter{line}
        \plabelx{#1} \theline \>}

\newenvironment{pream}[1]%
{{\bf Protocol {#1}}\\[10pt]%
{\sf Messages}%
\dBa}%
{\dE}

\nc{\SE}{\end{equation*}}
\nc{\SB}{\begin{equation*}}

\newcommand{\start}[1]{\ls{0.3}%
\renewcommand{\pname}{#1}%
\daddcontentsline{lot}{table}{\debug{\fbox{\pname}}}%
\begin{minipage}{\textwidtha}%
\setcounter{protblock}{0}%
\begin{pream}{#1}}
\newcommand{\starta}[1]{\ls{0.3}%
\begin{minipage}{\textwidtha}%
\begin{algo}{#1}}

\newcommand{\startz}{\ls{0.3}%
\begin{minipage}{\textwidtha}%
\begin{algoz}}
\newcommand{\startpa}[2]{\ls{0.3}
\renewcommand{\pname}{#1}%
\daddcontentsline{lot}{table}{\debug{\fbox{\pname}}}%
\begin{minipage}{\textwidtha}%
\setcounter{protblock}{0}%
{\bf Protocol {#1}}\\[10pt]%
\begin{algo}{#2}}
\newcommand{\aEz}{\end{algoz}}
\nc{\aE}{\end{algo}}
\newcommand{\mE}{\end{minipage}\ls{1.2}}
\newcommand{\pE}{\end{pream}}
\nc{\mB}{\begin{minipage}}

\newenvironment{algoz}%
{\ls{0.3}
\begin{tabbing}%
0000\debug{000} \= 111 \= 222 \= 333 \= 444 \= 555 \= 666 \= 777
\= 888 \= 999 \= 000 \kill
\prntvar\\%
}%
{\end{tabbing}}
\newenvironment{algo}[1]%
{\ls{0.3}\setcounter{line}{0}%
\begin{tabbing}%
0000\debug{000} \= 111 \= 222 \= 333 \= 444 \= 555 \= 666 \= 777
\= 888 \= 999 \= 000 \kill
\prntvar\\%
{\em Algorithm for node {#1}}\\}%
{\end{tabbing}}
{\ls{0.3}\setcounter{line}{0}%
\begin{tabbing}%
0000\debug{000} \= 111 \= 222 \= 333 \= 444 \= 555 \= 666 \= 777
\= 888 \= 999 \= 000 \kill
\prntvar\\%
{\em {#1}}\\}%
{\end{tabbing}}
{\begin{figure*}[t]
\begin{center}
\begin{minipage}{\textwidth}
\begin{tabbing}
0000\debug{11} \= 111 \= 211 \= 311 \= 411 \= 555 \= 666 \= 777 \kill }%
{\end{tabbing}
\end{minipage}
\caption[]{  \bf Design of a \ffi for $i$.}
\dlabel{designF}
\end{center}
\end{figure*}}

\newcommand{\prntvar}%
{\verb + itemsep = + \the\itemsep%
\verb + leftmargin = + \the\leftmargin%
\verb + leftskip = + \the\leftskip%
\verb + parindent = + \the\parindent%
\verb + leftmargini = + \the\leftmargini}

\renewcommand{\prntvar}{}

\newcommand{\FB}{\begin{figure}[hbtp]\centering}
\newcommand{\FE}[3]{\caption{#3%
	\debug{\fbox{\sname #2}}\debug{\fbox{#1}}}%
	\label{\sname #2} \end{figure}}
\newcommand{\tB}{\begin{table}[hbtp]\centering}
\newcommand{\tE}[3]{\caption{#3
	\debug{\fbox{\sname #2}}\debug{\fbox{#1}}}%
	\label{\sname #2}\end{table}}

\newcommand{\msec}[2]{\renewcommand{\sname}{}\section[#1
	\debug{\fbox {#2}}]{#1 \cmd{msec} \dlabelx{#2}}%
	\markboth{\today}{Sec. \thesection}}
\newcommand{\msubsection}[2]{\subsection[#1 \debug{\fbox {#2}}]
	{#1 \cmd{msubsection} \dlabelx{#2}}%
	 \markboth{\today}{Sec. \thesection}}

\newtheorem{definition}{Definition}
\newtheorem{notation}{Notation}

\newtheorem{theorem}{Theorem}
\newtheorem{lemma}{Lemma}

\renewcommand{\dB}{\begin{description}}

\input{tcilatex}

\renewcommand{\dB}{\begin{description}}
\nc{\mathbb}{\textbf}
\nc{\centerdot}{\cdot}

\begin{document}

\pagenumbering{gobble}
\clearpage
\thispagestyle{empty}

\title{
HDR - A Hysteresis-Driven Routing Algorithm for Energy Harvesting Tag Networks\footnote{Thanks are due to Maria Gorlatova and Gil Zussman for many discussions in the initial stages of this research and to Alexander Lavzin for useful comments.} \\
}
\author{Adrian Segall}

\maketitle

\begin{abstract}
The work contains a first attempt to treat the problem of routing in networks with energy harvesting units.  We propose HDR - a Hysteresis Based Routing Algorithm and analyse it in a simple diamond network.  We also consider a network with three forwarding nodes.  The results are used to give insight into its application in general topology networks and to general harvesting patterns.
\end{abstract}

\newpage
\clearpage
\tableofcontents
\clearpage
\pagenumbering{arabic}

\newpage

\newcommand{\fname}{consensus.tex}

\msec{Introduction}{Introduction}

Recent advances in the design of ultra-low power transceivers and solar cells has made it possible to develop and implement networks with self sustainable energy harvesting devices (\emph{EnHANTs}) that communicate with neighboring devices over wireless links (see e.g. \dcite{GWZ}, \dcite{GMSplus} and references therein).  In such networks, node energy increases via harvesting, in addition to it being spent by data transmission and reception.  As a result, the algorithms in \emph{EnHANTs} networks differ considerably from the ones in legacy sensor and ad-hoc networks.  Moreover, with the devices we are considering, the available energy for control and processing is extremely low and thus the employed algorithms must be simple and the amount of transmitted control data must be minimized.

The present work is the first attempt to design routing protocols for \emph{EnHANTs} networks.  We shall analyse the performance of a routing protocol for a very simple network.  Then we indicate how this simple analysis can be employed to provide insight into the design of routing algorithms in larger networks.

\msec{The Model}{Model}

\begin{figure}[hbtp]
\begin{center}
\includegraphics[scale=1]{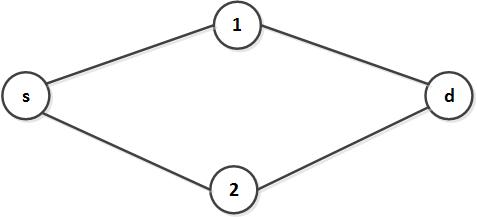}
\caption{The Model\debug{\fbox{fig-Model}}\label{fig-Model}}
\end{center}
\end{figure}

Consider a diamond network with 4 nodes as in Fig. \dref{fig-Model}.  Nodes $s,d$ are respectively the source and destination nodes.  Node $s$ generates data to be transferred via the energy-harvesting nodes 1 and 2 to the destination $d$.
Data is included in packets of fixed size. Time is divided in slots, where all nodes can change activity only at the end of a slot. Therefore, throughout this paper, we shall interchangeably use the terms \emph{"time"} and \emph{"end of slot"}.  Each wireless transmission by the source is overheard by both forwarding nodes, but in any given slot only one of them, referred to as \emph{the active node} and denoted by $v$, is forwarding the packets to the destination.  Only the active node spends energy in receiving the packet and forwarding it.  The other node is said to be \emph{inactive} and is denoted by $x$.  If node 1,2 is \emph{active}, we also say that we use \emph{route} 1,2 respectively.

At the end of each slot, the intermediate nodes inform the destination node of their current energy levels.  The active node can do this by piggy-backing the information to the last data packet in the given slot, whereas the inactive node needs to use a control message, referred to as a \emph{status message}.  Based on this information, the destination node decides what route to use and informs the forwarding nodes accordingly.  The decision is sent by the destination node just after having received the status messages, in a control message referred to as a \emph{switch-command message}.  We assume that it takes negligible time, but not negligible energy, to send the status messages, to receive the switch command and to apply it by the forwarding nodes.

For the first part of this work we assume that the energy harvesting rates at the two forwarding nodes is constant over a sufficiently long time, at the rate of $e_1,\,e_2$ $mJ / slot$ respectively.  We also assume for most of this work that the source generates and transmits data at a fixed rate of $g\,\, packets/slot$.  The source and the destination are assumed to be energy unlimited.
If we momentarily disregard the energy spent by control messages, the maximal data rate that can be transmitted from source to destination in this network is
\SB
g_{max} = (e_1 + e_2)/c\text{   ;    } packets/slot
\SE
where $c$ is the total combined energy required to receive and send a data packet.  This is consistent with the results in many works that treat conditions for maximum flow, for example \dcite{MSZ}, as applied to this simple network.  In order to achieve the maximal flow, the traffic must be split between the two routes at average rates $e_1/c$ and $e_2/c$ respectively.  Obviously, measurement of the harvesting rates and splitting the traffic accordingly are not easy tasks.  The purpose of this work is to design a simple routing protocol that hopefully achieves this split without the need to measure the harvesting rates.  We may mention in addition that if this goal is indeed achieved, the split will automatically adapt to (slow) changes in the harvesting rates.

The proposed algorithm, referred to as the \emph{Hysteresis-Driven Routing(HDR) Algorithm} is as follows.  The destination assigns as \emph{active} the node with the \emph{higher energy level}, except that, in order to avoid fast oscillations and high control overhead, it switches routes only when the energy level at the inactive node exceeds the level at the active node by a certain \emph{threshold}.  The threshold for switching in one direction may be different from the one in the opposite direction.  Since activity can change only at the end of a slot, if the threshold is reached during a slot, the \emph{actual switch time is at the end of that slot}.

\msubsection{Notations and assumptions}{notations}

For simplicity, we shall take the transmission energy for each status message to be the same, whether the information is piggybacked on a data message or sent separately.  We need to distinguish between the time \emph{just before} the transfer of the control messages and the time \emph{just after}.  We shall refer to the former as \emph{the time just before the end of the slot} and to the latter as \emph{the time at the end of the slot}.

We shall use the following notations.  The period between two consecutive switches in the same direction is referred to as a \emph{cycle}.

\begin{description}
\item{$e_u$} - energy harvested by node $u$ in each slot  ($mJ/$slot)
\item{$g$} - number of data packets generated by the source node in each slot (packets/slot)
\item{$c$} = combined energy to receive and transmit a data packet ($mJ/$packet)
\item{$c_t$} = energy to transmit a status control packet or status information piggybacked on a data packet($mJ$)
\item{$c_r$} = energy to receive a switch command packet ($mJ$)
\item{$B^{-}_u(i)$} - battery level at node $u$ \emph{just before} the end of slot $i$ in units of energy($mJ$)
\item{$B_u(i)$} - battery level at node $u$ \emph{at} the end of slot $i$, in units of energy ($mJ$)
\item{$B_{max}$} - maximum battery level ($mJ$)
\item{$B_{min}$} = $c_t + c_r$ - minimum battery level that allows transmission of control messages ($mJ$)
\item{$h_1$} - threshold to switch from route 1 to route 2 ($mJ$)
\item{$h_2$} - threshold to switch from route 2 to route 1 ($mJ$)
\item {$h$} = $h_1+h_2$ ($mJ$)
\item {$\gamma^c_u$} = \emph{node cycle throughput} = number of data packets transferred via node $u$ in each cycle (packets/cycle)
\item {$\gamma^c$} = \emph{total cycle throughput} = $\gamma^c_1 + \gamma^c_2$
\item{$1_z$} = 1 if $z$ is true and $0$ otherwise
\end{description}

\msec{Energy equations}{energy}

In the sequel, we develop formulas for the battery level at each node just before the end of each slot and at the end of each slot.  We also determine the maximal steady state flow  $g_s$ of data.
Since the harvesting rates are not known and they may, hopefully slowly, change with time, we shall also examine the behaviour of the system for arbitrary input rates $g$, both larger and smaller than $g_s$.
In order to avoid the treatment of too many cases however, we shall restrict the input rate to be larger than $max(e_1 ,e_2)/c$.  In fact, if for example, $c\,g < e_1 - c_t$, then the energy of node 1 always increases, even when node 1 is active; not a very interesting scenario.  We shall also assume that $min(e_1, e_2) > c_t + c_r$.

With the above assumptions, when the battery levels are away from the boundaries $B_{min}$ and $B_{max}$, the battery level at the inactive node $x$ increases from \emph{the end} of a slot until \emph{just before} the end of the next slot by $e_x\,\,mJ$.  The battery level at the active node $v$ decreases from \emph{the end} of some slot until \emph{just before the end} of the next slot by $(c\,g - e_v)\,\,mJ$.  If the battery at the inactive node reaches $B_{max}$, its harvesting stops.  For input rate $g$, when the battery level at the active node $v$ is away from $B_{min}$, it transfers $g$ packets/slot.

It remains to consider the activity of the active node when it is close to the boundary $B_{min}$.  If the active node battery reaches $B_{min}$ \emph{during} a slot, say a fraction $\alpha\,$ of the slot period into the slot time, then it will transfer $\alpha g $ packets before it reaches $B_{min}$  and  $(1-\alpha)\,e_v/c$ packets afterwards\footnote{For simplicity of notation and analysis, we disregard the fact that we only send whole packets.  The simulation does take this fact into account.}.
Next consider the situation at the end of some slot $i$.  Recall the order of transmissions starting just before the end of a slot time: first the status messages and then, if necessary, the switch command.  If just before the end of the slot, node $v$ is at $B^{-}_v(i) < B_{min}$,  it may run into a problem.  If we allow it to send the status message, thereby spending $c_t \,\, mJ$ for control, it will not have sufficient energy to receive the switch command, if any.  Thus in order to be safe, in this case we instruct the node \emph{to refrain from sending the status message} at the end of slot $i$.  The destination will assume that the node is at $B_{min}$ when it does not hear the status message and will act accordingly\footnote{Here, as well as throughout this work, we assume no lost messages.}.
Finally, consider data transmission.  If after sending the status message, the energy is less than $B_{min}$, we do not allow transfer of data messages in the next slot, with the hope that the harvested energy will sufficiently increase the battery level to allow it to send the status message at the end of the next slot.  Otherwise, the normal algorithm applies.

With the above considerations, we can write down the dynamics of the system.

\ls{0.3}%
\renewcommand{\pname}{B}%
\daddcontentsline{lot}{table}{\debug{\fbox{\pname}}}%

\begin{minipage}{\textwidtha}%
\ls{0.3}\setcounter{line}{0}%
\begin{tabbing}%
0000\debug{000} \= 111 \= 222 \= 333 \= 444 \= 555 \= 666 \= 777
\= 888 \= 999 \= 000 \kill
\prntvar\\%
\\%
\stepcounter{protblock}
\nl{bc}\>switch = 0
\nl{aa}\>$\bif B^{-}_x(i) - B^{-}_v(i) \geq h_v $
\nl{ab}\2   switch = 1
\nl{bx}\>$\emph{\textbf{end}}$
\nl{ac}\>$B_x(i)= B^{-}_x(i) - c_t$
\nl{m}\> $\bif (B^{-}_v(i) \geq B_{min})$
\nl{a}\>\>		$B_v(i) = B^{-}_v(i)  - c_t$
\nl{d}\> $\belse$		
\nl{e}\>\>      $B_v(i) = B^{-}_v(i)$
\nl{f}\> $\emph{\textbf{end}}$
\nl{y}\>$\bif$ (switch == 1)
\nl{b}\2        $B_v(i) = B_v(i) - c_r\text {   ;    }B_x(i) = B_x(i) - c_r $
\nl{g}\2        $v$ and $x$ switch
\nl{ag}\2       $B^{-}_x(i+1) = B_x(i) + e_x $
\nl{ad}\2       $B^{-}_v(i+1) = B_v(i) + e_v - c\,g$
\nl{ae}\2       $\gamma^c_v(i+1) = \gamma^c_v(i) +  g$
\nl{x}\>$\belse$
\nl{ax}\2       $B^{-}_x(i+1) = min(B_x(i) + e_x, B_{max}) $
\nl{n}\2        $\bif  (B_v(i) < B_{min})$
\nl{j}\3            $B^{-}_v(i+1) = B_v(i) + e_v$
\nl{i}\3            $\gamma^c_v(i+1) = \gamma^c_v(i) $
\nl{p}\2        $\belse$
\nl{k}\3            $B^{-}_v(i+1) = max(B_{min},B_v(i) +e_v -c\,g)$
\nl{ak}\3           $\alpha = max(0,min(1,(B_v(i)-B_{min})/(c\,g - e_v)))$
\nl{u}\3            $\gamma^c_v(i+1) = \gamma^c_v(i) + \alpha \, g + (1-\alpha)\,e_v/c$
\nl{w}\2        $\emph{\textbf{end}}$
\nl{z}\>$\emph{\textbf{end}}$
\end{tabbing}
\end{minipage}
\ls{1.2}

It is easy to see that the energy just after the transmission of the status control message, if any, is no smaller than $c_r$, so there is always energy to receive the switch message (see lines \dref{Bb} and \dref{Bab} in the Code above).  To avoid confusion, we point out that the code above is not performed by any node.  It merely describes the dynamics of the system.

Assume that we have just switched at a slot, referred to as slot $0$, from route 2 to route 1.
This happens because the threshold $h_2$ is reached during slot $0$ or just before its end and was not reached during the previous slot, namely
\EB
h_2 + e_1 - e_2 + c\,g > B^{-}_1(0) -  B^{-}_2(0) \geq h_2
\EE{eq-sw1}
The condition that we switch again $I_1$ slots afterwards is
\EB
h_1 + e_2 - e_1 + c\,g >  B^{-}_2(I_1) -  B^{-}_1(I_1) \geq h_1
\EE{eq-caps}

The differences in battery levels are calculated as sums of the expressions in lines \dref{Ba}, \dref{Be}, \dref{Bb}, \dref{Bk} in the Code above.
We have a similar condition for the next switch from route 2 to route 1.

\msec{Operation away from the boundaries}{away2}

Denote by
\dB
\item{$I_1^a$} = number of slots when node 1 is active in a given cycle $a$, when the system is away from the boundaries
\item{$I_2^a$} = number of slots when node 2 is active in a given cycle $a$, when the system is away from the boundaries
\item{$I^a$} = $I_1^a + I_2^a$
\item{$\gamma$} = $\gamma^c(a) / I^a$ = average throughput per slot in cycle $a$ (packets/slot)
\dE

When the system operates away from the boundaries, all \emph{max} and \emph{min} operands in the Code above do not apply.  Thus we have
\EB
h_1 + e_2 - e_1 + c\,g > B^{-}_2(I_1^a) - B^{-}_1(I_1^a) = I_1^a\,\, e_2    + I_1^a\,(c\,g - e_1)\, + B^{-}_2(0) - B^{-}_1(0) \geq h_1
\EE{eq-Iaax}
or
\EB
I_1^a = \lceil\frac{(h_1 + B^{-}_1(0) - B^{-}_2(0))}{(c\,g -e_1 + e_2)}\rceil
\EE{eq-Iaa}
where $\lceil\,Z\rceil$ denotes \emph{the smallest integer larger than or equal to $Z$}.
Similarly
\EB
I_2^a = \lceil\frac{(h_2 + B^{-}_2(I_1^a) - B^{-}_1(I_1^a))}{(c\,g -e_2 + e_1)}\rceil
\EE{eq-Iaay}

We shall be interested in the drift in total energy of the system $\Delta_{total} = B_1(I^a) - B_1(0) + B_2(I^a) - B_2(0)$ during a cycle $I^a = I_1^a + I_2^a$.   The energy harvested during a cycle is $(e_1 + e_2)\,I^a$.  Since $g$ packets are sent in each slot, the energy spent is $c\,g\,I^a$.  At the end of each slot, each node spends $c_t$ for control.  There are 2 switch times in a cycle and at every switch time each node spends $c_r$.  The total drift in energy of both nodes per cycle is given by the difference between the harvested energy and the spent one:
\EB
\Delta_{total} = -4\,c_r + (e_1 + e_2 - 2\,c_t -c\,g)\,I^a \,\,mJ/cycle
\EE{eq-BDtotal}

This is how far we can get analytically with arbitrary parameters.  A useful approximation is to select thresholds as multiples of the net slot energy change.  This implies that the thresholds are reached exactly at slot time, a fact that considerably simplifies the analysis .

\msec{Thresholds divide slot energies - Operation away from the boundaries}{divide}

Suppose a switch from route 2 to route 1 occurs at the end of slot $0$ and at that time, the energy level difference exactly matches the threshold value.  This means $B^{-}_2(0) - B^{-}_1(0) = h_2$.  Then, with $h = h_1 + h_2$, we have
\EB
I_1^a=\lceil\frac{h}{ c\,g+(e_2-e_1)}\rceil
\EE{eq-I1aaa}
and a similar expression for $I_2^a$.
If $h$ divides evenly the slot energies $c\,g+(e_2-e_1)$ and $c\,g+(e_1-e_2)$, then
\EB
I_1^a=\frac{h}{ c\,g+(e_2-e_1)}\text{     ;     }
I_2^a=\frac{h}{ c\,g+(e_1-e_2)}
\EE{eq-Ia}
\EB
I^a = \frac{2c\,g\,h}{(c\,g)^2 - (e_2 - e_1)^2}
\EE{eq-away}
Moreover, the battery level difference returns to its initial value at the end of a cycle, namely
$B^{-}_2(I^a) - B^{-}_1(I^a) = B^{-}_2(0) - B^{-}_1(0) = h_2$.

In each slot $g$ packets are transferred, thus the average throughput is
\EB
\gamma = g \text{      packets/slot}
\EE{throughput}
independently of the thresholds.
The split ratio between the two routes is
\EB
\frac{\gamma_1^c}{\gamma_2^c} = \frac{g\,I_1^a}{g\,I_2^a} = \frac{c\,g+e_1-e_2}{c\,g+e_2-e_1}
\EE{eq-split0}

In the sequel, we shall also need the following quantities.
Let $\delta B_u (I)$ denote the change in the battery level at node $u$ during an interval of $I$ slots.  Then
\EB
\delta B_1(I_1^a) = -\frac{h\,(c\,g -e_1)}{ c\,g+(e_2-e_1)} \text{        ;         }
\delta B_2(I_1^a) = \frac{h\, e_2}{ c\,g+(e_2-e_1)}
\EE{eq-Delta}
and
\EB
\delta B_1(I_2^a) = \frac{h\, e_1}{ c\,g+(e_1-e_2)} \text{        ;         }
\delta B_2(I_2^a) = -\frac{h(c\, g - e_2)}{ c\,g+(e_1-e_2)}
\EE{eq-Delta1}

The drift $\Delta  = B_1(I^a) - B_1(0) = B_2(I^a) - B_2(0)$ is half the total drift (\dref{eq-BDtotal}), namely
\EB
\Delta = -2\,c_r   + \frac{(e_1 + e_2 -2\,c_t-c\,g)\,I^a}{2}\,\,mJ/cycle
\EE{eq-B}

\msec{Negligible Control - Thresholds divide slot energies}{negli}

If the energy spent for sending and receiving control messages is negligible, we have
$c_t = c_r = B_{min} = 0$.

\msubsection{Operation away from the boundaries}{awayq}

For negligible control, the drift is $\Delta = (e_1 + e_2 - c\,g)\,I^a/2$.
Let $g_s$ denote the input that induces steady state, namely $\Delta = 0 $.  We get $g_s = (e_1 + e_2)/c$.  If $g = g_s$, then the condition for the switches to occur at exactly battery level differences $h_1$ and $h_2$ becomes that $h/(2\,e_1)$ and $h/(2\,e_2)$ are integers.  For example for $e_1 = 0.8\, , e_2 = 0.6$, any two threshold values $h_1, h_2$ that add up to 4.8 or its multiples will do the job.

We conclude that if $g = g_s$, then the \emph{HDR Algorithm} indeed transfers in dynamic steady state the entire input rate $g_s$.   For other values of the input rate $g$, the behavior is transient:  the battery levels will drift up if $g < g_s$ until they reach steady state close to $B_{max}$ and down if $g > g_s$ until they reach steady state close to $B_{min}$.  The steady state behavior in these cases is treated in the following sections.

\subsection{Drifts and Operation close to boundaries}\dlabel{close}

As before, the analysis here is for negligible control and thresholds evenly dividing the slot energies.  Since in practice, the harvesting rates vary, it is important to investigate the behavior of the system  when $e_1 + e_2$ is not necessarily equal to $c\,g$.

If $ e_1 + e_2 > c\,g$, then the battery levels drift up, until at least one of them reaches $B_{max}$.  Similarly, if $ e_1 + e_2  < c\,g$, the drift is down, until at least one of the batteries reaches the lower bound $B_{min} = 0$.  Before reaching the boundary, the switch times are as before Eq. (\dref{eq-Ia}).  The drift per slot is $\Delta/I^a$, with $\Delta$ and $I^a$ given in Eq. (\dref{eq-B}) and (\dref{eq-away}).

When a node is \emph{inactive}, its battery charges. Recall that we have assumed $c \,g > e_1, c\,g > e_2$, so the battery level at the \emph{active} node goes down.  Thus, only the inactive node can reach the boundary $B_{max}$ and only the active node can reach the lower boundary $B_{min} = 0$.  We assume that $B_{max}$ is much larger than the thresholds, so that if one of the nodes reaches some boundary, the other does not reach the opposite boundary in the same cycle.

In addition to the assumption of switching with differences in battery levels equal to the thresholds, in the analysis below we shall also assume that boundaries are reached by nodes exactly at slot times.  As before, we point out that the simulation is performed without these assumptions.

Take $0$ to indicate the time of a switch when node $v$ becomes \emph{actiVe}.  Recall that we denote by  $x$ the other node, namely the one that will become active \emph{neXt}.  The condition for the next switch (from route $v$ to route $x$) to occur after slot $I_v$ is given by Eq. (\dref{eq-caps}).  Note that the switch time $I_v$ when boundaries are reached may not be the same as $I_v^a$, the switch time when no boundary is met.
If neither node reaches a boundary (0 or $B_{max}$), then the switch time is as before $I_v=I_v^a$.

Let
\begin{description}
  \item[$I'_v$] = slot at the end of which node $v$ reaches 0 before a switch, where $0 < I'_v \leq I_v^a$
  \item[$I''_v$] = slot at the end of which node $x$ reaches $B_{max}$ before a switch, where $0 < I''_v \leq I_v^a$
\end{description}
If the corresponding boundary is not reached before a switch, then $I'_v \, , I''_v$ respectively is defined to be equal to $I_v$.  Note that if either node reaches some boundary before a switch, it stays at that boundary until the next switch, and the other node cannot reach the opposite boundary.  Also note that until a boundary is reached, both the $min$ and the $max$ operands in the Code above are inactive.
The switch condition (\dref{eq-caps}) becomes:
\EB
e_x\,\cdot I''_v + (\, c\, g\, - e_v)\,I'_v = h
\EE{eq-swCC}

The cycle dynamics is given below.

\ls{0.3}%
\renewcommand{\pname}{X}%
\daddcontentsline{lot}{table}{\debug{\fbox{\pname}}}%

\begin{minipage}{\textwidtha}%
\ls{0.3}\setcounter{line}{0}%
\begin{tabbing}%
0000\debug{000} \= 111 \= 222 \= 333 \= 444 \= 555 \= 666 \= 777
\= 888 \= 999 \= 000 \kill
\prntvar\\%
{\em Node v is actiVe, node x is neXt} \` /* $x = v \, mod \, 2 + 1$ \\%
\stepcounter{protblock}
\nl{m}\> $I_v^a = h/(c\,g + e_x -e_v)$
\nl{s}\>$\bcase (B_{max}-B_x(0))/e_x < I_v^a$ \{   \` /*node $x$ reaches boundary $B_{max}$
\nl{a}\>\>		$I''_v = (B_{max}-B_x(0))/e_x$
\nl{x}\>\>		$I_v = (h - (B_{max} - B_x(0)))/(c\,g - e_v)$
\nl{r}\>\>      $B_v(I_v) = B_v(0) - I_v\,(c\,g - e_v)\text{    ;     }B_x(I_v) = B_{max}$
\nl{f}\>\>      $\gamma^c_v=I_v \, g$ \\
	\>\>	\}
\nl{y}\>$\bcase B_v(0)/(c\,g-e_v) < I_v^a$\{   \`/*node $v$ reaches boundary $0$
\nl{b}\2        $I'_v = B_v(0)/(c\,g-e_v)$
\nl{g}\2        $I_v = (h - (B_v(0)))/e_x$
\nl{c}\2        $B_v(I_v) = 0\text{    ;     }B_x(I_v) = B_x(0) + I _v\,e_x$
\nl{h}\2        $\gamma^c_v=I'_v\, g+ (I_v-I'_v)\,(e_v)/c$ \\
    \2       \}
\nl{i}\> $\belse$\{      \` neither node reaches boundary
\nl{j}\3            $I_v = I_v^a$
\nl{p}\3            $B_v(I_v) = B_v(0)-I_v\,(c\,g - e_v)\text{    ;    }B_x(I_v) = B_v(0)+I_v\,c\,g$
\nl{k}\3            $\gamma^c_v = I_v \, g$ \\
    \2       \} \\

\end{tabbing}
\end{minipage}
\ls{1.2}

\msubsection{System Throughput and Operation close to the boundaries}{low}

In this Section we examine the operation of the system in terms of throughput and split ratio, at low and high battery level.

\begin{figure}[hbtp]
\begin{center}
\includegraphics[scale=0.5]{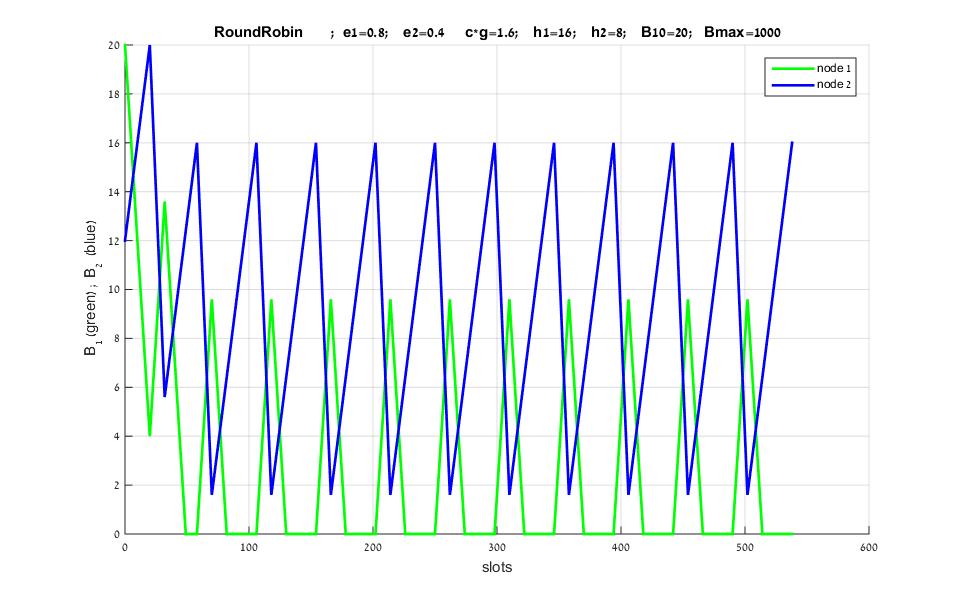}
\caption{Only node 1 is at boundary 0 for a non-zero period of time\debug{\fbox{fig-case1}}\label{fig-case1}}
\end{center}
\end{figure}

We start with low level.  If $e_1 + e_2 < c\,g$, the battery levels will drift down while seesawing.  An example appears in Fig. \dref{fig-case1}.

As we shall see in the sequel, after the time when one of the nodes reaches the boundary 0, the system operates in a periodic dynamic steady state.  There are two cases: i) only one node is ever empty for a non-zero period of time ; ii) both nodes alternately reach 0.  The condition for these to happen is given below.

\LB{lemma-0}
If $h_1/h_2 \ge (c\,g - e_2)/e_1$, then $B_2$ is never empty for a non-zero period of time and the throughput during a cycle, the cycle length, and the average throughput are as in Eq. (\dref{eq-cycle}) below.
\LE

\PB
The condition above is equivalent to $h_1\, e_1 + h_2\,e_2 - h_2 \,c\,g \ge 0$.
The switch from route 1 to route 2 occurs after slot $I$
when $B_2(I) = B_1(I) + h_1 \ge h_1$.
Afterwards, $B_2$ goes down at a rate $(c\, g -e_2)\,\mu J /slot$.
Node $B_2$ will reach 0 only if during the next $I_2^a$ slots (namely before the next switch) it will do so.
Thus if $h_1-I_2^a\cdot(c\,g - e_2)  \ge 0$, the battery $B_2$ will never be at $B_{min}$ for a non-zero period.
Substituting for $I_2^a$ from Eq. (\dref{eq-away}), we obtain the condition above.

Since the system drifts down, at least one battery must hit 0 and if the condition in the Lemma holds, only $B_1$ can do so.  After the first time it hits 0, it waits for $B_2$ to reach $h_1$, at which time a switch from route 1 to route 2 occurs.  Rename the slot at the end of which this occurs as $I=0$.  From this slot on, the system is in dynamic steady state.

The next switch, from 2 to 1, occurs at time $I_2^a$ , and the battery levels are
\EB
B_1(I_2^a) = I_2^a\cdot e_1=\frac{e_1\,h}{c\,g + (e_1-e_2)}
\EE{eq-bat1}
and
\EB
B_2(I_2^a) = B_2(0)+ \delta B_2(I_2^a) =
h_1-\frac{(h_1+h_2)(c\,g-e_2)}{c\,g +(e_1-e_2)} =
\frac{h_1\,e_1 - h_2\,(c\,g -e_2)}{c\,g +(e_1-e_2)}
\EE{eq-bat2}
If we again rename the slot after which the switch from route 2 to route 1 occurs as $I=0$, the battery $B_1$ drains out at time
\SB
I'_1=  \frac{e_1\,h}{(c\,g + (e_1-e_2))(c\,g - e_1)}
\SE
and then node 1 waits for $B_2$ to charge to level $h_1$ at time
\SB
(h-\frac{e_1\,h}{c\,g + (e_1-e_2)})/e_2=\frac{h(c\,g-e_2)}{e_2 \cdot (c\,g+(e_1-e_2))}\text{   ,  }
\SE
where the switch from 1 to 2 occurs at battery levels (\dref{eq-bat1}) , (\dref{eq-bat2}).  The scenario repeats from now on.

While active and in dynamic steady state, node 2 transfers a total of
$\gamma^c_2 = (g\,h)/(e_1 - e_2 + c\,g)\text{    packets/cycle } $ , and node 1 transfers
$\gamma^c_1 = (e_1\, g\,h)/ (e_2\,(e_1 - e_2 + c\,g))\text{    packets/cycle } $.  The split ratio is
\SB
\gamma_1/\gamma_2 =e_1/e_2
\SE
The total throughput in a cycle $\gamma^c$, the number of slots in a cycle $I$ and the average throughput $\gamma = \gamma^c/I$ are:
\EB
\gamma^c = \frac{g\,(e_1+e_2)\,h}{e_2\,(e_1-e_2+c\,g)}\text {   ;   }
I = \frac{c\,g\,h}{e_2\cdot(c\,g + e_1 - e_2)}\text{   ;    }
\gamma = \frac{e_1+e_2}{c} \text{     packets/slot}
\EE{eq-cycle}
Note that, since $c\,g > e_1 + e_2$, the average throughput above is the maximal possible.
$\otimes$

\LB{lemma-2}
If $h_1/h_2 \leq e_2/(c\,g - e_1)$  , then $B_1$ never empties out and the throughput during a cycle, the cycle length and the average throughput are as in Eq. (\dref{eq-cycle}) with indexes 1 and 2 interchanged.
\LE

\LB{lemma-1}
If  $ e_2/(c\,g - e_1) < h_1/h_2 < (c\,g-e_2)/e_1$ , then both batteries alternately reach level 0 and stay there for non-zero numbers of slots, and the throughput during a cycle, the cycle length and the average throughput are as in Eq. (\dref{eq-cycle1}).
\LE

\begin{figure}[hbtp]
\begin{center}
\includegraphics[scale=0.5]{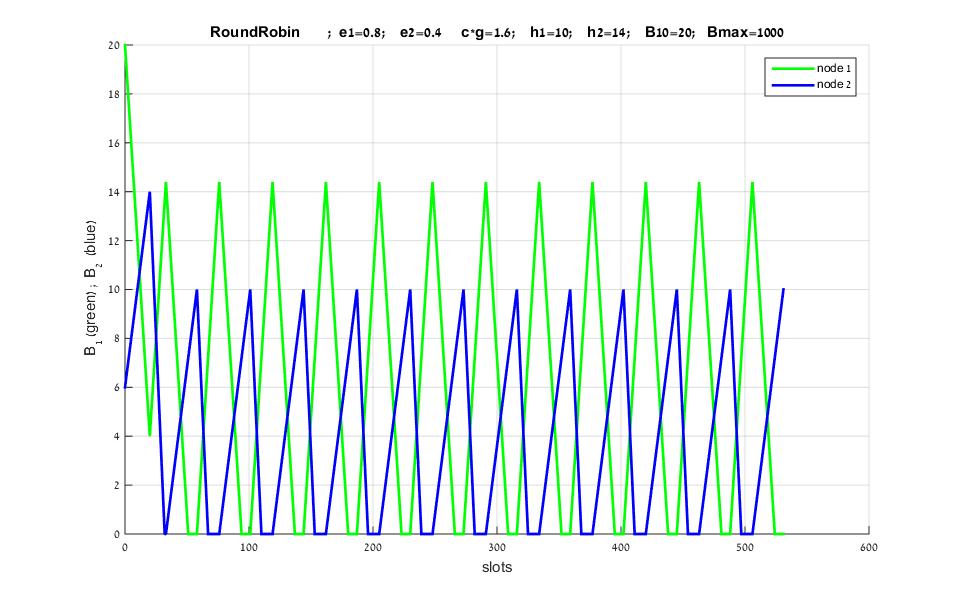}
\caption{Both nodes empty out alternately\debug{\fbox{fig-case2}}\label{fig-case2}}
\end{center}
\end{figure}

\PB
An example appears in Fig. \dref{fig-case2}.
Since the system drifts down, at least one battery must eventually drain out.  Suppose node 1 is the \textbf{first} that hits battery level 0 at a time \emph{that is not its lower tip} .  Then it waits for $B_2$ to reach $h_1$, at which time a switch from route 1 to route 2 occurs.  Let $I=0$ indicate the slot after which this happens.  From this time on, the system is in dynamic steady state.

At time $I'_2 = h_1/(c\,g -e_2)<h_2/e_1$, the battery $B_2$ reaches 0.  It waits for $B_1$ to reach $h_2$ at time $I_2= h_2/e_1$, at which time the switch from node 2 to node 1 occurs (see Fig. \dref{fig-case2}).  The battery levels at this time are
\EB
B_1(I_2) = h_2\text{      ;      } B_2(I_2) =  0
\EE{eq-bat3}
Counting slots after that switch, $B_1$ drains out at time $I'_1=  h_2/(c\,g - e_1)<h_1/e_2$, waits for $B_2$ to charge to level $h_1$ at time $h_1/e_2$, where the switch from 1 to 2 occurs at battery levels corresponding to (\dref{eq-bat3}) with  exchanged indices.

While active, node 2 transfers a total of
$\gamma^c_2 = (e_1\,h_1+e_2\,h_2)/(c\,e_1)$
packets, and node 1 transfers
$\gamma^c_1 = (e_1\,h_1+e_2\,h_2)/(c\,e_2)$.  The split ratio is
$\gamma^c_1/\gamma^c_2 =e_1/e_2$.
The total throughput in a cycle $\gamma^c$, the total number of slots in a cycle $I$ and the average throughput $\gamma= \gamma^c/I$ are
\EB
\gamma^c = \frac{(e_1+e_2)(e_1\,h_1 + e_2\,h_2)}{c\,e_1\,e_2}\text{    ;    }
I = \frac{h_1\,e_1+h_2\,e_2}{e_1 \,e_2}\text{    ;    }
\gamma  =  \frac{e_1 + e_2}{c} \text{     packets/slot}
\EE{eq-cycle1}
$\otimes$

\begin{figure}[hbtp]
\begin{center}
\includegraphics[scale=0.5]{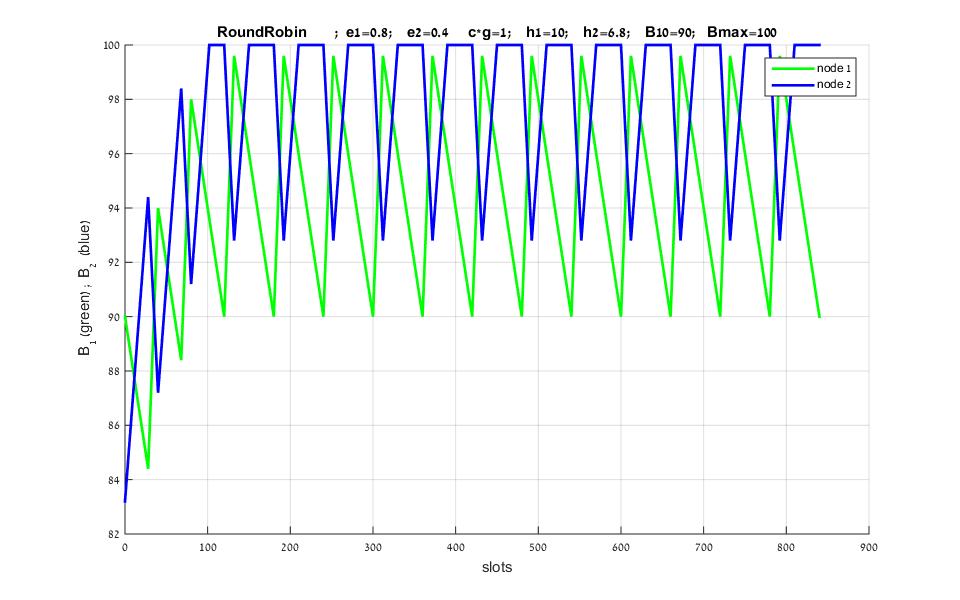}
\caption{Node 2 reaches $B_{max}$\debug{\fbox{fig-case3}}\label{fig-case3}}
\end{center}
\end{figure}

Next we consider the case when the input rate is less that the total harvesting rates, namely $e_1 + e_2 > c\,g $.  In this case the battery levels will drift up while seesawing.  The operation here is similar to the one at low-level batteries. An example when only node 2 reaches $B_{max}$ appears in Fig. \dref{fig-case3}.  The properties for all cases are stated in the summary section below without proofs.  The proofs are similar to the ones above.

\msubsection{Summary of Analytic Results}{summary}

\TB{T-Summary}

Under the assumptions that:
\eB
\item the energies $c_t,c_r$ spent for control  are negligible
\item all switches occur at times when the difference in battery levels is exactly equal to the thresholds
\item boundaries are reached at slot time
\eE
the system behaves as follows
\eB
\item
While away from the boundaries
\SB
I_1=I_1^a =\frac{h}{c\,g + e_2-e_1}\text{  slots  ;    }
I_2=I_2^a=\frac{h}{c\,g + e_1-e_2}\text{     slots }
\SE
\SB
\gamma^c_1 = \frac{g\,h}{c\,g + e_2-e_1}\text{ packets   ;    }
\gamma^c_2 = \frac{g\,h}{c\,g + e_1-e_2}\text{    packets}
\SE
\SB
\frac{\gamma^c_1}{\gamma^c_2} = \frac{c\,g + e_1-e_2}{c\,g + e_2-e_1} \text{    ;    }
\gamma^c = \frac{2\,c\,g^2\,h}{(c\,g)^2 - (e_1-e_2)^2 }\text{ packets   ;   }
\SE
\SB
I = \frac{2\,c\,g\,h}{(c\,g)^2 - (e_1-e_2)^2 } \text{   slots   ;   }
\gamma = g\text{    packets/slot  ; }
\text{Drift per cycle} = \frac{e_1 + e_2 - c\,g}{2}\,\,\mu J
\SE
\item
If $ e_1+e_2 = c\,g$ and the battery levels are initially far from the boundaries, then they are in dynamic steady state with the following:
\SB
I_1=I_1^a =\frac{h}{2\,e_2}\text{  slots  ;    }
I_2=I_2^a=\frac{h}{2\,e_1}\text{     slots }
\SE
\SB
\gamma^c_1 = \frac{g\,h}{2\,e_2}\text{ packets   ;    }
\gamma^c_2 = \frac{g\,h}{2\,e_1}\text{    packets}
\SE
\SB
\frac{\gamma^c_1}{\gamma^c_2} = \frac{e_1}{e_2} \text{    ;    }
\gamma^c = \frac{c\,g^2\,h}{2\,e_1\,e_2}\text{ packets   ;   }
\SE
\SB
I = \frac{c\,g\,h}{2\,e_1\,e_2} \text{   slots   ;   }
\gamma = g\text{    packets/slot  ; }
\text{Drift per cycle} = 0
\SE
\item
If $ e_1+e_2 < c\,g$, then the battery levels will drift down while seesawing and
\dB
\item{(a)} If $h_1/h_2 \le e_2/(c\,g - e_1)$  , then $B_1$ is never at level zero level for a non-zero amount of time and in steady state:
\SB
I_1 = \frac{h}{c\,g + (e_2-e_1)}\text{  slots   ;    }
I_2= \frac{h(c\,g-e_1)}{e_1 \cdot (c\,g+(e_2-e_1))}\text{  slots  ;   }
\SE
\SB
\gamma^c_1 = \frac{g\,h}{e_2 - e_1 + c\,g}\text{   packets ;   }
\gamma^c_2 = \frac{e_2\, g\,h}{e_1\,(e_2 - e_1 + c\,g)}\text{    packets  ;   }
\SE
\SB
\frac{\gamma^c_1}{\gamma^c_2} =\frac{e_1}{e_2}{      ;      }
\gamma^c = \frac{g\,(e_2+e_1)\,h}{e_1\,(e_2-e_1+c\,g)}\text {   ;   }
I = \frac{c\,g\,h}{e_1\cdot(c\,g + e_2 - e_1)}\text{   ;    }
\gamma = \frac{e_1+e_2}{c} \text{     packets/slot}
\SE
\item{(b)}
If  $h_1/h_2 \ge (c\,g - e_2)/e_1$, then $B_2$ is never at level zero for a non-zero amount of time and in steady state (see Eq. (\dref{eq-cycle})):
\SB
I_1= \frac{h(c\,g-e_2)}{e_2 \cdot (c\,g+(e_1-e_2))}\text{    ;   }
I_2 = \frac{h}{c\,g + (e_1-e_2)}\text{    ;   }
\gamma^c_1 = \frac{e_1\, g\,h}{e_2\,(e_1 - e_2 + c\,g)}\text{    ;   }
\gamma^c_2 = \frac{g\,h}{e_1 - e_2 + c\,g}
\SE
\SB
\frac{\gamma^c_1}{\gamma^c_2} =\frac{e_1}{e_2}\text{     ;     }
\gamma^c = \frac{g\,(e_1+e_2)\,h}{e_2\,(e_1-e_2+c\,g)}\text {   ;   }
I = \frac{c\,g\,h}{e_2\cdot(c\,g + e_1 - e_2)}\text{   ;    }
\gamma = \frac{e_1+e_2}{c} \text{     packets/slot}
\SE
\item{(c)}
If  $ e_2/(c\,g - e_1) < h_1/h_2 < (c\,g - e_2)/e_1$, then in steady state both batteries alternately reach level 0 for non-zero amounts of time and (see Eq. (\dref{eq-cycle1})):
\SB
I_1 = \frac{h(c\,g - e_2)}{e_2 ( c\,g +e_1 - e_2)}\text {   ;   }
I_2 = \frac{h}{c\,g +e_1-e_2}\text {   ;   }
\gamma^c_2 = \frac{e_1\,h_1+e_2\,h_2}{c\,e_1}\text {   ;   }
\gamma^c_1 = \frac{e_1\,h_1+e_2\,h_2}{c\,e_2}
\SE
\SB
\frac{\gamma^c_1}{\gamma^c_2} =\frac{e_1}{e_2}\text{    ;    }
\gamma^c = \frac{(e_1+e_2)(e_1\,h_1 + e_2\,h_2)}{c\,e_1\,e_2}{    ;    }
I = \frac{h_1\,e_1+h_2\,e_2}{e_1 \,e_2}\text{    ;    }
\gamma  =  \frac{e_1 + e_2}{c} \text{     packets/slot}
\SE
\dE
\item
If $e_1+e_2 > c\,g$, then the battery levels will drift up while seesawing\footnote{In a previous version of this report, the conditions below were wrongly stated.  Now they are correct - thanks to Alex Lavzin for perceiving the error.}
\dB
\item{(a)} If $h_1/h_2 \ge e_1/(c\,g-e_2)$ , then $B_1$ is never at $B_{max}$ for a non-zero amount of time and in steady state:
    \SB
    I_1 = \frac{e_1 \,h }{(c\,g - e_1)(c\,g +e_1-e_2)}\text{    ;    }
    I_2 = \frac{h}{c\,g +e_1 - e_2}\text{    ;    }
    \gamma^c_1 = \frac{e_1 \,h\,g }{(c\,g - e_1)(c\,g +e_1-e_2)}\text{    ;    }
    \SE
    \SB
    \gamma^c_2 = \frac{h\,g}{c\,g +e_1 - e_2}\text{    ;   }
    \gamma^c = \frac{ c\, g^2\, h }{ (c\,g - e_1) (c\,g + e_1 - e_2 )}\text{    ;   }
    \frac{\gamma^c_1}{\gamma^c_2} = \frac{e_1}{c\,g - e_1}
    \SE
    \SB
    I = \frac{c\,g\,h }{(c\,g - e_1)(c\,g + e_1 -e_2)}\text{    ;        }
    \gamma = g
    \SE
\item{(b)} If $h_1/h_2 \le (c\,g-e_1)/e_2$ , then $B_2$ is never at $B_{max}$ for a non-zero amount of time and in steady state:
    \SB
    I_1 = \frac{h}{c\,g +e_2 - e_1}\text{    ;    }
    I_2 = \frac{e_2 \,h }{(c\,g - e_2)(c\,g +e_2-e_1)}\text{    ;    }
    \gamma^c_1 = \frac{h\,g}{c\,g +e_2 - e_1}
    \SE
    \SB
    \gamma^c_2 = \frac{e_2 \,h\,g }{(c\,g - e_2)(c\,g +e_2-e_1)}\text{    ;    }
    \gamma^c = \frac{ c\, g^2\, h }{ (c\,g - e_2) (c\,g + e_2 - e_1 )}\text{    ;   }
    \frac{\gamma^c_1}{\gamma^c_2} = \frac{c\,g - e_2}{e_2}
    \SE
    \SB
    I = \frac{c\,g\,h }{(c\,g - e_2)(c\,g + e_2 -e_1)}\text{    ;   }
    \gamma = g
    \SE
\item{(c)} If $(c\,g-e_1)/e_2 < h_1/h_2 < e_1/(c\,g-e_2)$  , then both $B_1$ and $B_2$ alternately reach $B_{max}$ and stay there for non-zero amounts of time and in steady state:
    \SB
    I_1 = \frac{h_1}{c\,g - e_1}\text{    ;   }
    I_2 = \frac{h_2}{c\,g - e_2}\text{    ;   }
    \gamma^c_1 = \frac{g\,h_1}{c\,g - e_1}\text{    ;   }
    \gamma^c_2 = \frac{g\,h_2}{c\,g - e_2}
    \SE
    \SB
    \gamma^c = \frac{  g\,(c\,g\,h -e_1\,h_2 - e_2\,h_1) }{ (c\,g - e_2) (c\,g - e_1 )}\text{    ;   }
    \frac{\gamma^c_1}{\gamma^c_2} = \frac{h_1\,(c\,g - e_2)}{h_2\,(c\,g - e_1)}\text{    ;   }
    I = \frac{  c\,g\,h -e_1\,h_2 - e_2\,h_1 }{ (c\,g - e_2) (c\,g - e_1 )}\text{    ;   }
    \gamma = g
    \SE
\dE

\eE
\TE
In words, we can summarize the dynamic steady state activity as follows:
\eB
\item If $e_1 + e_2= c\,g$, steady state occurs at all battery levels away from the boundaries. The throughput is $g$, with split ratio $e_1/e_2$, independent of the thresholds.  The thresholds affect only the frequency of switching.
\item If $e_1 + e_2 < c\,g$, steady state occurs close to empty battery levels.  The throughput is $(e_1 + e_2)/c$, with split ratio $e_1/e_2$, independent of the thresholds.  The thresholds affect only the frequency of switching.
\item If $e_1 + e_2 > c\,g$, steady state occurs close to full battery levels.  The throughput is $g$ with split ratio $\neq e_1/e_2$.  The thresholds affect the split ratio, as well as the frequency of switching.
\eE

\msec{Case when differences in battery levels at switch time do not exactly match the thresholds}{relax}

The switch from route 1 to route 2 (and viceversa) occurs at the end of the slot during which $h_1$ (respectively $h_2$) is reached.  If $h_1$ or $h_2$ is reached during a slot (as opposed to just before the end of the slot), the difference in battery levels at switch time is larger than the corresponding threshold.

Consider the case $e_1 + e_2 = c\,g$, namely steady state when away from the boundaries.   If the system starts with $B_2 = B_1 + h_2$ and if $h = h_1 + h_2$ is a multiple of both $2\,e_1$ and $2\,e_2$, then all switches will occur with battery level differences equal to the thresholds. Moreover, the battery levels at the end of each cycle are the same as the ones at the beginning of the cycle.  Namely the dynamic steady state has \emph{a period of one cycle}.

If $h$ is not a multiple of  $2\,e_1$ and $2\,e_2$, then thresholds will be reached not necessarily at the end of slots.  As a result, the difference in battery levels at switch time will be larger than the threshold values.  On the other hand, since $e_1 + e_2 = c\,g$, the battery levels do not drift and thus the system must be in dynamic steady state.  The behavior is now that the battery levels do not return to their initial values after each cycle, but rather after more than one cycle.  As a topic for future research, it will be interesting to find conditions for periods containing 1,2,3,.. cycles.  We present in Table \dref{Table-Patterns} a few numerical examples for $e_1 = 0.6$, $e_2 = 0.8$, $c= 0.08$, $g = 17.5$.  The third column is the number of cycles in each period.  The first row in each entry are the switch times, namely the slots at the end of which the switch occurs. The second row represents the battery level differences at switch time.

\begin{table}[hbtp]
\begin{center}
\begin{tabular}{|l|l|c|c|c|c|c|c|c|c|}
  \hline
 $h_1$  &$h_2$ & period & \multicolumn{7}{c|}{switch at slot ;  $B(1)- B(2)$}\\ \hline
   4  &0.8 & 1 & 0   & 3    & 7  & & & &\\
      &    &   & 0.8 & -4 & 0.8  & &  & &\\ \hline
  2   & 1  & 1 & 0   & 3    & 7  & & & &\\
      &    &   & 1.4 & -3.4 & 1.4  & &  & &\\ \hline
  6.2 & 5  & 2 &  0  &  7   & 17  & 25  & 35 & & \\
      &    &   &  5  & -6.2 & 5.8 & -7 & 5 & &  \\ \hline
  5   & 5  & 3 &  0  & 7    & 17  & 24 & 33 &40 &49 \\
      &    &   &  5  & -6.2 & 5.8 & -5.4 & 5.4 & -5.8& 5\\
  \hline
\end{tabular}
\caption{Switching Patterns \debug{\fbox{Table-Patterns}}\label{Table-Patterns}}
\end{center}
\end{table}

\msec{Non-negligible Power Consumption for Control Messages}{nonnegligible}

Here we consider the situation with non-zero $c_t$ and $c_r$ and look only at the operation away from the boundaries.  If $h$ evenly divides the slot energies, the drift $\Delta$ and the cycle length are as in \dref{eq-B}, (\dref{eq-away})

\begin{figure}[hbtp]
\begin{center}
\includegraphics[scale=0.5]{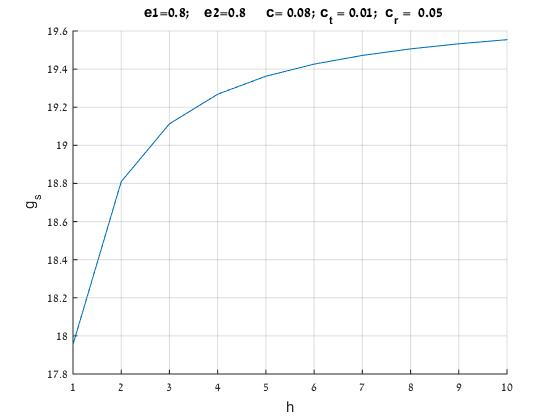}
\caption{$g_a$ as a function of $h$\debug{\fbox{fig-g}}\label{fig-g}}
\end{center}
\end{figure}

Substituting $I^a$ into $\Delta$, we get
\EB
\Delta = -2\,c_r + \frac{c\,g\,h\,(\tilde{e} - c\,g)}{(c\,g)^2 - (e_2 - e_1)^2}\,\,mJ/cycle
\EE{eq-BD1}
where $\tilde{e} = e_1 + e_2 -2\,c_t$.
The condition for dynamic steady state is $\Delta = 0$.  We obtain a quadratic equation for $g$, whose solution, denoted by $g_s$, is given by
\EB
g_s = \frac{h\,\tilde{e}+\sqrt{h^2\,\tilde{e}^2 + 8\,c_r\,(2\,c_r+h)\,(e_2 - e_1)^2}}{2\,c\,(2\,c_r+h)}
\EE{eq-g}
The throughput $g_s$ as a function of the thresholds is shown in Fig. \dref{fig-g}.  As expected, the higher the thresholds the less energy is spent on control, since the inter-switch period goes up, and thus more energy is left for data messages.  However, high thresholds can lead to the batteries reaching the boundary.  With the values of $e_1,\,e_2,\,c$ as in the graph, the value of $g_s$ for negligible control is $(e_1+e_2)/c = 20$.

\begin{figure}[hbtp]
\begin{center}
\includegraphics[scale=0.5]{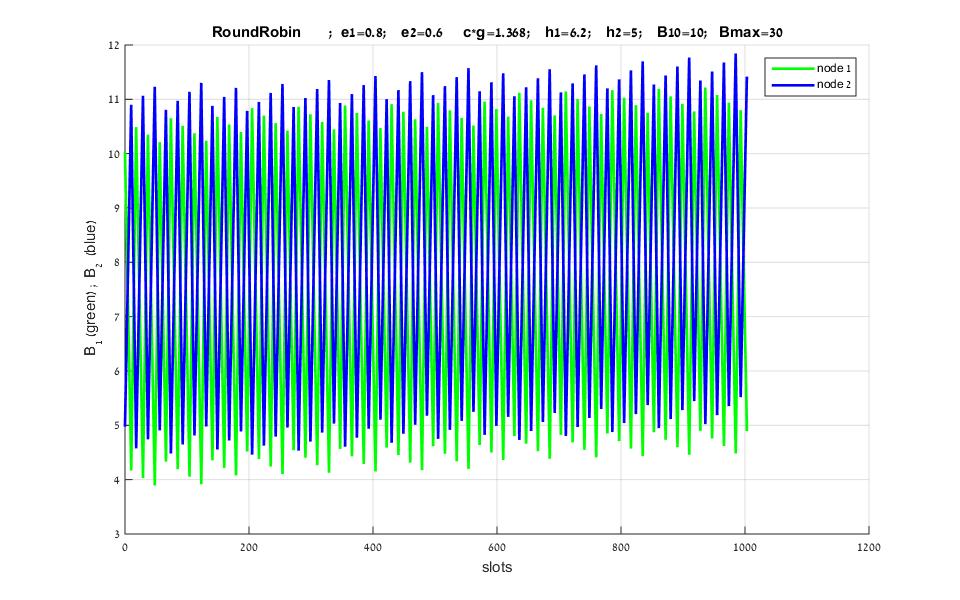}
\caption{Behavior of system with $g = g_s$ \debug{\fbox{fig-crct}}\label{fig-crct}}
\end{center}
\end{figure}

\begin{figure}[hbtp]
\begin{center}
\includegraphics[scale=0.5]{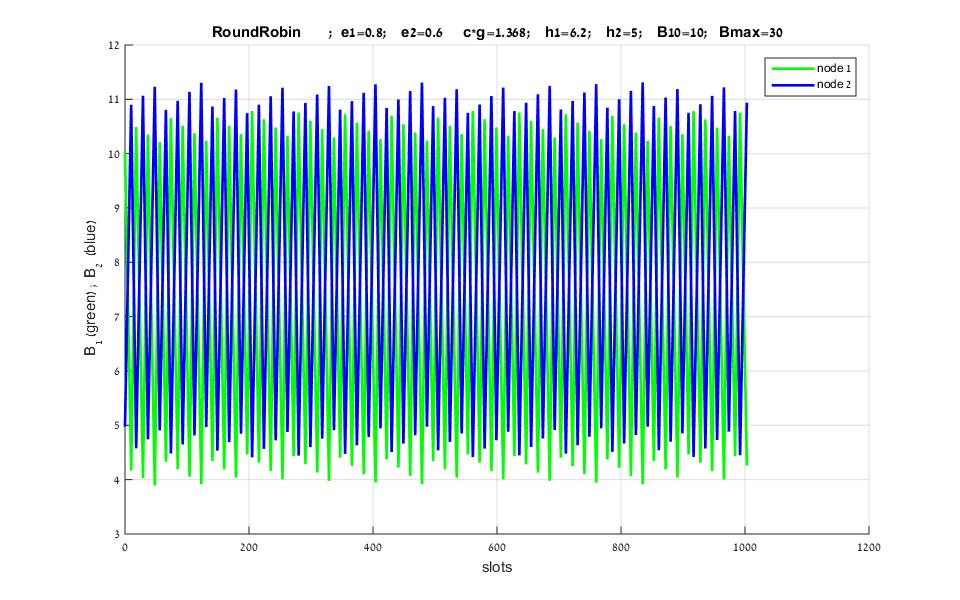}
\caption{Behavior of system with $g$ obtained by estimating $I^a$  \debug{\fbox{fig-crctFeedback}}\label{fig-crctFeedback}}
\end{center}
\end{figure}

At this point, recall that the analysis is correct if $I_1^a$ and $I_2^a$ are integers.
It is difficult to get explicit conditions for this to hold for arbitrary control values $c_t$ and $c_r$ and we have to resort to numerical simulations.
Consider for example the situation in Fig. \dref{fig-crct}, where $e_1 = 0.8\,\,;\,\,e_2 = 0.6\,\,;\,\,c=0.08\,\,;\,\,h_1 = 6.2\,\,;\,\,h_2 = 5\,\,;\,\,c_r = 0.05\,\,;\,\,c_t = 0.01$.
We get $g_s = 17.10$.  The plot shows a drift of about $0.5\, mJ$ in 1000 slots.
This is due to the fact that an average cycle length turns out to be 18.72 slots, as opposed to the theoretical 16.73 slots given by Eq. (\dref{eq-away}).
As an exercise, we have performed the same simulation, but at each step we estimate the current average cycle length and adjust the input rate $g$ so that $\Delta$ given by Eq. (\dref{eq-B}) is zero.
Not surprisingly, with this feedback, there is no drift (see Fig.\dref{fig-crctFeedback}).

\msec{Summary so far}{Implications}

The analysis above is performed for a simple system with several helpful assumptions.  We have extended it in several directions, like non-negligible power spent on control messages, but we can look at its implications to more general practical systems.  In practice, the harvesting rates are time varying, according to various parameters, like time of day and illumination conditions.  Our discussion here assumes that those variations are relatively slow. If the pattern is known, the source can try to adapt to the current parameters.  For example, it should increase the input rate $g$ when the harvesting rates increase.  Also, in this simple network, the wireless transmissions of the intermediate nodes are overheard at the source.  The latter can use this information in order to adapt the input rate.

Consider now extensions of the \emph{HDR Algorithm} to larger networks, either with one data collection point (several sources, one destination) or with several source-destination pairs.  Many routing procedures have been proposed for legacy ad-hoc and sensor networks with a variety of performance criteria, like maximum lifetime or maximum total throughput (see e.g. \dcite{AY}, \dcite{ASSC}, \dcite{KK}, \dcite{LS} and references therein).  In such networks the units spend energy and thus the battery level only goes down.  Those algorithms do not seem to be applicable to networks with energy harvesting units, where energy also increases via harvesting.  In networks with harvesting nodes, lifetime and total throughput have no meaning.

The analysis in this first work on the topic of routing in harvesting node networks can provide an insight for larger networks.  Assuming that the topology of the network does not change often (e.g. tags on books in a library, static tags in a room or building), two or more paths can be established in advance for every source-destination pair.  If all units have similar harvesting rates, it makes sense to select node-disjoint paths.  If there are units with significantly larger harvesting rates than the others, they can participate in more than one path.  One can think of several simple centralized or distributed algorithms to select the paths, but this topic is outside the scope of the present work.  The energy levels of the nodes on the path can be periodically collected, either piggy-backed on data messages or via control messages.   Using a threshold mechanism as given by the \emph{HDR Algorithm} on the maximum battery level along the path, the destination can decide which path to use.  Again, the exact procedures for collecting data and for informing nodes in the network upon the path selection are topics for future research.  For instance, one can also consider situations when the collection node has a powerful transmitter that can be simultaneously heard by all nodes.  In this case, it can directly communicate routes and switching decisions to all nodes.

\msec{Three Forwarding Nodes}{three}

\begin{figure}[hbtp]
\begin{center}
\includegraphics[scale=1]{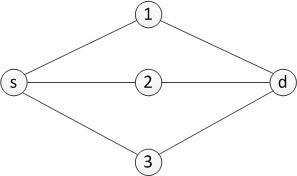}
\caption{Three Forwarding Nodes\debug{\fbox{fig-3nodes}}\label{fig-3nodes}}
\end{center}
\end{figure}

Having extensively discussed the diamond configuration and having obtained detailed results for its behavior, we can now adventure into looking at larger networks.
In this section we analyze the behaviour of the \emph{HDR Algorithm} in a network with three parallel paths (see Fig. \dref{fig-3nodes}).

We first consider a simple switching policy, whereby the routes are switched in a round-robin fashion, with thresholds $h_1$, $h_2$, $h_3$ for the switches (1,2),(2,3),(3,1) respectively.

\msubsection{Operation away from the boundaries}{away3}

Let $v$ denote the current actiVe node and $x$ denote the neXt active node.  Let $I_v^a$ denote the number of slots when the destination employs route $v$ while the batteries \emph{are away from the boundaries}.  We also assume as before that switching occurs at battery level differences exactly matching the value of the thresholds.

If we denote $I^a = I_1^a + I_2^a + I_3^a$ and refer to $I^a$ as a \emph{cycle}, then the operation away from the boundaries is:
\begin{eqnarray}
\nonumber B^{-}_1(0)=B{-}_3(0 )+h_3\text{     ;     }B_u(0) = B^{-}_u(0) - c_t -c_r\\
\nonumber B^{-}_2 (I_1^a) = B^{-}_1(I_1^a) + h_1\text{     ;     }B_u(I_1^a) = B^{-}_u(I_1^a) - c_t -c_r\\
\nonumber B^{-}_3 (I_1^a+I_2^a) = B^{-}_2(I_1^a+I_2^a) + h_2\text{     ;     }B_u(I^a) = B^{-}_u(I^a) - c_t -c_r\\
\nonumber B_1(I_1^a) = B_1(0) - (c\,g+c_t -e_1)\,I_1^a - c_r \\
\nonumber B_2(I_1^a) = B_2(0) + (e_2-c_t)\, I_1^a - c_r \\
\nonumber B_3(I_1^a) = B_3(0) + (e_3-c_t)\, I_1^a -c_r\\
\nonumber B_1(I_1^a+I_2^a) = B_1(I_1^a) + (e_1-c_t)\,I_2^a - c_r \\
\nonumber B_2(I_1^a+I_2^a) = B_2(I_1^a) - (c\,g + c_t - e_2)\,I_2^a -c_r \\
\nonumber B_3(I_1^a+I_2^a) = B_3(I_1^a) + (e_3 - c_t)\,I_2^a -c_r\\
\nonumber B_1(I^a) = B_1(I_1^a+I_2^a) + (e_1-c_t)\,I_3^a - c_r\\
\nonumber B_2(I^a) = B_2(I_1^a+I_2^a) + (e_2 - c_t)\,I_3^a - c_r\\
B_3(I^a) = B_3(I_1^a+I_2^a) - (c\,g + c_t - e_3)\,I_3^a - c_r
\end{eqnarray}
We also require that we are in dynamic steady state, namely that the change in battery level during a cycle is the same for all three batteries (denoted by $\Delta$):
\SB
\Delta = B_1(I^a)-B_1(0) = B_2(I^a)-B_2(0) = B_3(I^a) - B_3(0)
\SE
Substituting above, and selecting the initial condition $B_1(0) = B10$, the system of linear equations has a unique solution.  We give here explicitly only the expressions for $I_1^a$, $I_2^a$, $I_3^a$ and for the drift.
\begin{eqnarray}
\nonumber I_1^a = \frac{h\,(c\,g + 2\,e_1 -e_2 -e_3)}{2\,(c^2\,g^2 -e_1^2 - e_2^2 -e_3^2 +e_1\,e_2+e_2\,e_3 +e_3\,e_1)} \\
\nonumber I_2^a = \frac{h\,(c\,g + 2\,e_2 -e_1 -e_3)}{2\,(c^2\,g^2 -e_1^2 - e_2^2 -e_3^2 +e_1\,e_2+e_2\,e_3 +e_3\,e_1)} \\
I_3^a = \frac{h\,(c\,g + 2\,e_3 -e_1 -e_2)}{2\,(c^2\,g^2 -e_1^2 - e_2^2 -e_3^2 +e_1\,e_2+e_2\,e_3 +e_3\,e_1)} \dlabel{eq-away3}
\end{eqnarray}

The throughput is $g$ packets/slot and the split ratio is
\SB
\gamma_1 : \gamma_2 : \gamma_3 = (c\,g + 2\,e_1 -e_2 -e_3) : (c\,g + 2\,e_2 -e_1 -e_3) : (c\,g + 2\,e_3 -e_1 -e_2)
\SE

The cycle length and the drift are
\SB
I^a = \frac{3\,c\,g\,(h_1 + h_2 + h_3)}{2\,(c^2\,g^2 -e_1^2 - e_2^2 -e_3^2 +e_1\,e_2+e_2\,e_3 +e_2\,e_3)}
\SE
\EB
\Delta = -3c_r - \frac{c\,g\,h\,(c\,g - \tilde{e})}{2\,(c^2\,g^2 -E )}
\EE{eq-3drift}
where $\tilde{e} = e_1 + e_2 + e_3-3\,c_t$ and $E = e_1^2 + e_2^2 + e_3^2 - e_1\,e_2 - e_2\,e_3 -e_3\,e_1$.

The system is in steady state if $\Delta = 0$, which occurs at input rate $g = g_s$ given by the solution of the corresponding quadratic equation
\EB
g_s = \frac{h\,\tilde{e}+\sqrt{h^2\,\tilde{e}^2 + 24\,c_r\,(6\,c_r+h)\,E}}{2\,c\,(6\,c_r+h)}
\EE{eq-steady}

\begin{figure}[hbtp]
\begin{center}
\includegraphics[scale=0.5]{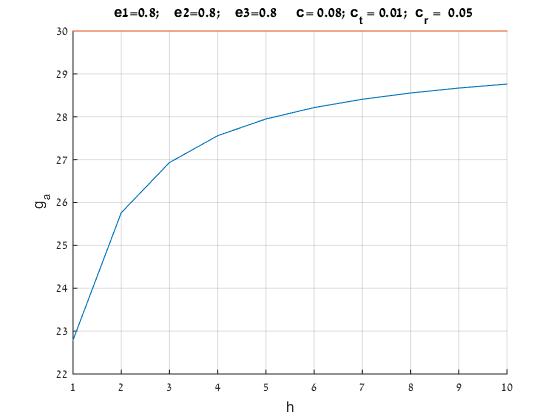}
\caption{$g_s$ as a function of $h$ for a system with 3 forwarding nodes\debug{\fbox{fig-g3}}\label{fig-g3}}
\end{center}
\end{figure}

For $c_r = c_t = 0$, the drift is zero if  $ e_1 +e_2 + e_3 = c\,g$, and then $ \gamma_1 : \gamma_2 : \gamma_3 = e_1 : e_2 : e_3$, namely there is no drift and the throughput through each node is proportional to its harvesting rate.

\msubsection{Operation close to the boundaries - negligible control}{close3}

Here we assume negligible energy spending for control messages.  If $e_1+e_2+e_3 > c\,g$, the battery levels drift up, until at least one of them reaches $B_{max}$.  Similarly, if $ e_1+e_2+e_3 < c\,g$, then the drift is down, until at least one of the batteries empties out.  Before reaching the boundary, the switch times are as before Eq. (\dref{eq-away3}).  The drift is $\Delta$.

Obviously, when a node is inactive, its battery charges. As before, we assume
$c \,g > max(e_1, e_2, e_3)$, so the battery level at the active node goes down.  Thus, only the inactive nodes can reach the boundary $B_{max}$ and only the active node can reach the boundary 0.

\emph{We look first at the system just after a switch from route 3 to route 1.}  We shall temporarily refer to the time when this occurs as slot 0.  The condition for the next switch (from route 1 to route 2) to occur at the end of slot $I_1$ is given by Eq. (\dref{eq-caps}), which here translates to

\EB
\sum\limits_{0<j\le I_1}( e_2\,1_{( B_2(j-1) < B_{max})}- e_1\,1_{(0< B_1(j-1) )} )+\sum\limits_{0<j\le I_1}c\,g\,1_{(0< B_1(j-1) )}=h_1+ B_1(0) - B_2(0)
\EE{eq-sw3}

Note that since the switch at time 0 is from node 3 and not from node 2, in general $B_1(0) - B_2(0) \neq h_2$.
Note that the switch time $I_1$ when boundaries may be reached for a non-zero amount of time may not be the same as $I_1^a$, the switch time when no boundary is met.

Since nodes 2 and 3 are inactive during $I_1$, their battery level increases, at rates $e_2 , \, e_3 \,\,mJ /slot$ respectively.  On the other hand, since $c\,g > e_1$, the battery level at the active node decreases, at a rate $(c\,g - e_1)\, mJ $/slot.  For now we assume that $B_{max}$ is large and hence only one of the boundaries can be reached in a given cycle.  If none of the three nodes reaches a boundary, then the switch time is at time
\SB
I_1^b = max\left(0,\frac{h_1 + B_1(0) - B_2(0)}{c\,g +e_2 -e_1}\right)
\SE
Let
\begin{description}
  \item[$I'_1$] = slot when node 1 reaches $0$ for a non-zero amount of time, where $0 < I'_1 \leq I_1^b$
  \item[$I''_1$] = slot when node 2 reaches $B_{max}$ for a non-zero amount of time, where $0 < I''_1 \leq I_1^b$
  \end{description}

Each of the two quantities above are defined to equal $I_1$ if the corresponding boundary is not reached for a non-zero amount of time.  Note that if any node reaches some boundary, it stays at that boundary until the next switch.
Therefore the switch condition (\dref{eq-sw3}) becomes:
\EB
e_2\cdot I''_1 + (\, c\, g\, - e_1\,)I'_1 = h_1 + B_1(0) - B_2(0)
\EE{eq-swCC3}
if $h_1 + B_1(0) - B_2(0) > 0$ and $I_1 = I'_1 = I''_1 = 0$ otherwise.
The  total number of packets transferred $\gamma^c_1$ is $I'_1\cdot g + (I_1-I'_1)\cdot (e_1/c) $ if $I'_1 < I_1^b$ and $I_1\cdot g $ otherwise.
The operation of the system is similar upon the switch from 2 to 3 and then from 3 to 1 (with a round robin substitution).

The evolution of the system is given below, where $B_1(0), B_2(0), B_3(0)$ are the battery levels at their respective time 0:

\vspace{1cm}

\ls{0.3}%
\renewcommand{\pname}{X3}%
\daddcontentsline{lot}{table}{\debug{\fbox{\pname}}}%

\begin{minipage}{\textwidtha}%
\ls{0.3}\setcounter{line}{0}%
\begin{tabbing}%
0000\debug{000} \= 111 \= 222 \= 333 \= 444 \= 555 \= 666 \= 777
\= 888 \= 999 \= 000 \kill
\prntvar\\%
{\em Node v is active, node x is next} \` /* $x = v \, mod \, 3 + 1$ \\%
\stepcounter{protblock}
\nl{m}\> $I^b = max(0,(h_v +B_v(0) - B_x(0)/(c\,g + e_x -e_v))$
\nl{s}\>$\bcase ((B_{max}-B_x(0))/e_x) < I^b$ \{   \` /*node $x$ reaches limit $B_{max}$
\nl{a}\>\>		$I''_v = (B_{max}-B_x(0))/e_x$
\nl{d}\>\>		$I_v = (h_v - (B_{max} - B_v(0))/(c\,g - e_v)$
\nl{e}\>\>      $B_v(I_v) = B_v(0)-I_v\,(c\,g-e_v)\text{   ;    }B_x(I_v) = B_{max}$
\nl{n}\>\>      $B_u(I_v) = min(B_u(0)+I_v\,e_u, B_{max})$\`/* $u$ is the third node
\nl{f}\>\>      $\gamma_v=I_v \, g$ \\
	\>\>	\}
\nl{y}\>$\bcase (B_v(0)/(c\,g-e_v) < I^b)$\{   \`/*node $v$ reaches limit 0
\nl{b}\2        $I'_v = B_v(0)/(c\,g-e_v)$
\nl{g}\3        $I_v = (h_v - B_x(0))/e_x$
\nl{r}\3        $B_v(I_v)=0\text{  ;   }B_x(I_v)= B_x(0)+I_v\,c\,g$
\nl{c}\3        $B_u(I_v)= B_u(0)+I_v\,e_u$\`/* $u$ is the third node
\nl{h}\3        $\gamma_v=I'_v\, g+ (I_v-I'_v)\,(e_v/c)$ \\
    \2       \}
\nl{i}\> $\belse$\{      \` neither node reaches limit
\nl{j}\3            $I_v = I^b$
\nl{p}\3            $B_v(I_v) = B_v(0)-I_v\,(c\,g - e_v)\text{    ;    }B_x(I_v) = B_v(0)+I_v\,c\,g$
\nl{o}\3            $B_u(I_v)= B_u(0)+I_v\,e_u$\`/* $u$ is the third node
\nl{k}\3            $\gamma_v = I_v \, g$ \\
    \2       \} \\

\end{tabbing}
\end{minipage}
\ls{1.2}

\msubsection{The Earliest Switch schedule}{ES}

In the Round Robin (RR) switching policy, the active routes are pre-assigned.  Suppose node 1 is active.  A switch will occur only when the battery level at node 2 exceeds the battery level at node 1 by the corresponding threshold.  On the other hand, the battery level at node 3 may exceed the corresponding threshold much earlier.  If we allow switches from 1 to 3 as well, the performance of the algorithm might improve.  To investigate this possibility, we consider here the \emph{Earliest Switch (ES) schedule}.  In \emph{ES}, routes are switched to the node that first reaches the corresponding threshold.  For simplicity, we look only at the case when the thresholds $h_{12}, h_{13}$ for the switch from 1 to 2 and from 1 to 3 respectively, are the same, namely $h_{12} = h_{13} = h_1$.  Similarly $h_{23} = h_{21} = h_2$ and
$h_{31} = h_{32} = h_3$.  Also, we assume negligible $c_t$ and $c_r$ and take $c = 0.08 mJ$/packet.

We have selected a series of sets of parameters as in Table \dref{Table-Param}.  We have run all simulations for a duration of 2000 slots, with statistics gathered only beginning at slot 301, to allow for the system to reach dynamic steady state.

\begin{table}[hbtp]
\begin{center}
\begin{tabular}{|c|c|c|c|c|c|c|c|c|l|}
\hline
Config & $c\,g$ & $e_1$   &   $e_2$   & $e_3$ & $h_1$ & $h_2$    & $h_3$ & Bmax & Comment \\ \hline
A &  1.6 & 0.1 & 0.7 & 0.8 & 5 & 10 & 10 & 100 &   $c\,g = e_1 +e_2 + e_3$ \\  \hline
B & 1.6 & 0.1 & 0.7 & 0.8 & 5 & 10 & 10 & 12 &    $c\,g = e_1 +e_2 + e_3$ \\  \hline
C & 2.4 & 0.1 & 0.7 & 0.8 & 5 & 10 & 10 & 100 &    $c\,g > e_1 +e_2 + e_3$ \\  \hline
D & 1.2 & 0.1 & 0.7 & 0.8 & 5 & 10 & 10 & 100 &     $c\,g < e_1 +e_2 + e_3$ \\  \hline
E & 0.2  & 0.1 & 0.01 & 0.01 & 5 & 10 & 10 & 60 &   $c\,g > e_1 +e_2 + e_3$ \\  \hline
F & 1.2 & 0.1 & 0.7 & 0.8 & 10 & 10 & 10 & 100 &     $c\,g < e_1 +e_2 + e_3$ \\  \hline
G & 1.6 & 0.1 & 0.7 & 0.8 & 5 & 10 & 10 & 60 &      $c\,g = e_1 +e_2 + e_3$ \\  \hline
\end{tabular}
\caption{Parameters\debug{\fbox{Table-Param}}\label{Table-Param}}
\end{center}
\end{table}

The results are given in the Table \dref{Table-Results}.  We could not find a conclusive statement as to when does the ES schedule perform significantly better then RR.  In general, when $Bmax$ is low, it seems that RR waists time in waiting for the corresponding threshold to be reached.

\begin{table}[hbtp]
\begin{center}
\begin{tabular}{|c|c|c|c|}
  \hline
Config  &                     & RR        & ES        \\ \hline
  A & Thruput                      & 20        & 20        \\ \cline{2-4}
    & ave cycle (slots)             & 32        & 34        \\ \cline{2-4}
    & $\#$ of switches             & 163       & 156        \\ \hline
  B & Thruput                      & 18.75     & 19.91      \\ \cline{2-4}
    & ave cycle (slots)             & 13.3      & 28.03      \\ \cline{2-4}
    & $\#$ of switches              & 163       & 157        \\  \hline
  C & Thruput                    & 20        & 20         \\ \cline{2-4}
    & ave cycle (slots)          & 46        & 18.52      \\ \cline{2-4}
    & $\#$ of switches           & 109       & 152        \\  \hline
  D & Thruput                       & 15        & 15        \\ \cline{2-4}
    & ave cycle (slots)          & 31.3      & 28.8      \\ \cline{2-4}
    & $\#$ of switches                & 100       & 91        \\  \hline
  E & Thruput                      & 20        & 20      \\ \cline{2-4}
    & ave cycle (slots)           & 32        & 34      \\ \cline{2-4}
    & $\#$ of switches              & 163       & 156        \\  \hline
  F & Thruput                     & 15        & 15      \\ \cline{2-4}
    & ave cycle (slots)          & 36        & 33      \\ \cline{2-4}
    & $\#$ of switches            & 82        & 85        \\  \hline
  G & Thruput                     & 20        & 20      \\ \cline{2-4}
    & ave cycle (slots)          & 30.8      & 33      \\ \cline{2-4}
    & $\#$ of switches            & 163       & 153        \\  \hline
\end{tabular}
\caption{Comparison of Throughputs\debug{\fbox{Table-Results}}\label{Table-Results}}
\end{center}
\end{table}

\begin{figure}[hbtp]
\begin{center}
\includegraphics[scale= 0.5]{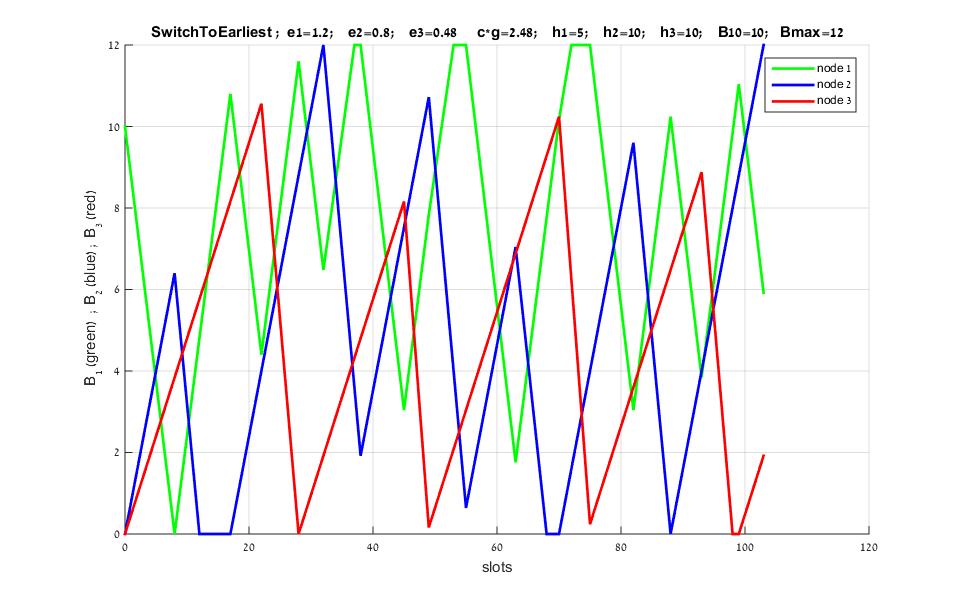}
\caption{Node 1 becomes active after node 2\debug{\fbox{fig-case6}}\label{fig-case6}}
\end{center}
\end{figure}

In the scenario of Fig. \dref{fig-case6}, node 2 becomes active starting at the end of slot 8 and its threshold is $h_2 = 10$. It reaches boundary 0 at end of slot 12 and waits for the other nodes to gather sufficient energy.  In RR (not depicted), it would wait for the battery level at node 3, its pre-assigned successor, to reach value 10, which would happen during slot 21.   However, the battery level at node 1 reaches value 10 earlier, during slot 17 and in ES, the destination will reroute to node 1 at the end of that slot.  The number of packets forwarded from end of slot 17 until the end of slot 21 is the following. In RR, node 2 is active at level 0 and thus it forwards data at rate $e_2/c = 10 \,\, packets/slot$, whereas in ES, node 1 is active away from boundaries and thus it transmits at rate $g = 31 \,\, packets/slot$, for a total gain of $(31-10) \cdot 4 = 84 \,\, packets$.

\msec{Variable Harvesting Rates and Inputs}{variableV}

\begin{figure}[hbtp]
\begin{center}
\includegraphics[scale=1]{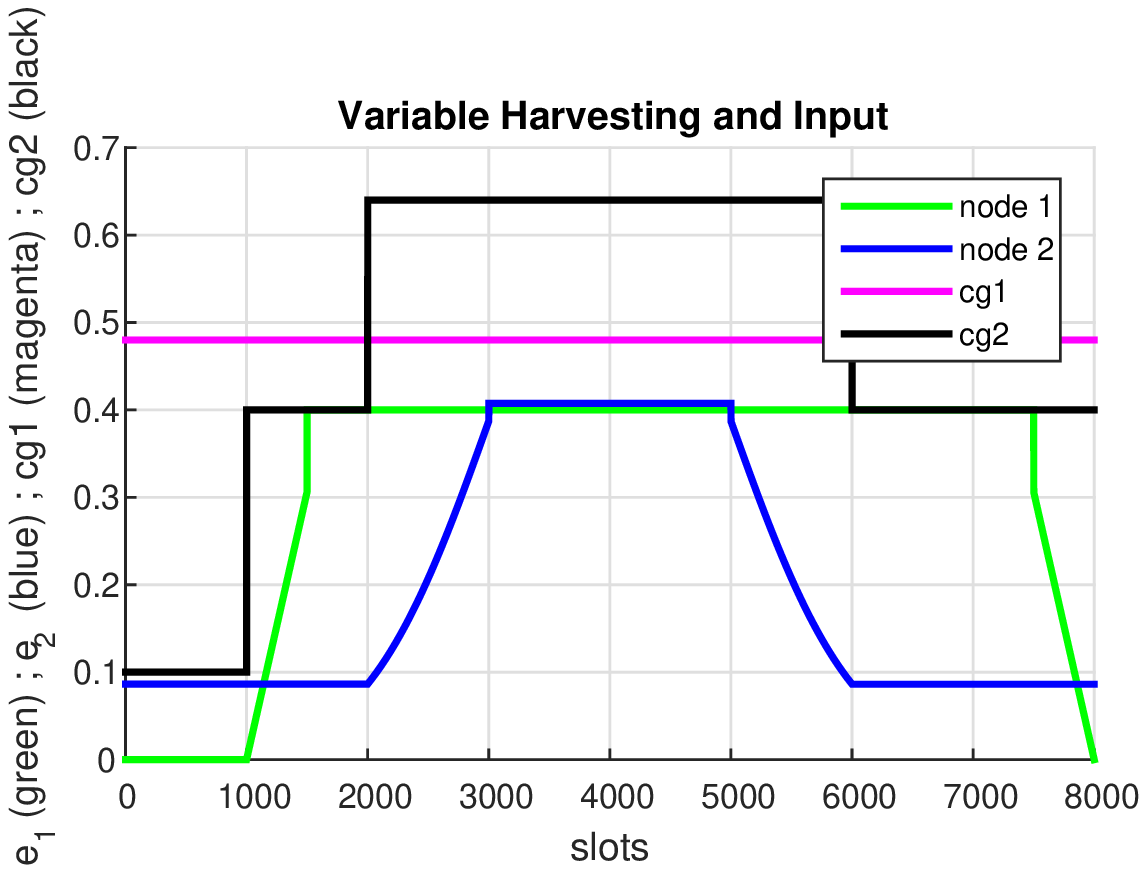}
\caption{Variable harvesting and input\debug{\fbox{fig-Harvest}}\label{fig-Harvest}}
\end{center}
\end{figure}

Practical systems cannot guarantee time invariant harvesting rates.  When harvesting rates and inputs are time varying, the energy equations are the same as in the Algorithm in Sec. \dref{energy}, except that the parameters $e_v(i),e_x(i),g(i)$ are time dependent.
Fig.\dref{fig-Harvest} approximates the pattern of harvesting of two nodes over 8,000 slots that appears in \dcite{GMSplus}.  We have investigated the behavior of the system with $c = 0.08, c_t = 0.01, c_r = 0.05, B_{max} = 200$ and two types of inputs, each totaling 48,000 packets over 8,000 slots.  With these parameters the total energy required to transfer the packets is 48,000 * 0.08 = 3,840 mJ.  During the period under consideration, the nodes harvest 2,553 and 1,595 mJ respectively, for a total of 4,148 mJ.  With transmission status messages requiring 160 mJ and receipt of switch commands requiring about 10 mJ, if the harvest and inputs were uniformly distributed, the harvest would have been more than sufficient to transfer all packets.

\begin{figure}[hbtp]
\begin{center}
\includegraphics[scale=0.35]{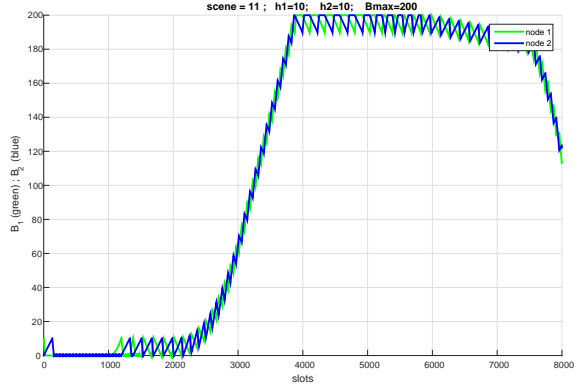}
\caption{Behavior of the System with Fixed Input $cg1/c$\debug{\fbox{fig-Variable1}}\label{fig-Variable1}}
\end{center}
\end{figure}

In our time-varying scenarios, the harvest changes with time.  In the first scenario, the input is uniformly distributed at level of $0.48/c$.  The contents of the batteries over the 8,000 slots is shown in Fig. \dref{fig-Variable1} and the inputs vs. the throughput over periods of 1000 slots each appears in Table \dref{Table-Thru1}.

\begin{table}[hbtp]
\begin{center}
\begin{tabular}{|c|c|c|c|c|c|c|c|c|}
  \hline
input &  6000 & 6000 &  6000 &  6000& 6000 & 6000  & 6000  & 6000  \\ \hline
throughput & 1072 & 4150 &  5983 & 5994 & 5994  & 5994 & 5994 & 5994  \\ \hline

\end{tabular}
\caption{Inputs vs. Throughput in Scenario 1\debug{\fbox{Table-Thru1}}\label{Table-Thru1}}
\end{center}
\end{table}

During the first 1000 slots, node 1 has no harvesting and, except for the first few slots where it uses the initial battery energy (10 mJ), it cannot transfer any packets.  Node 2 harvests 0.086 mJ/slot and thus can transfer roughly $g = e_2/c \cong 1 packet/slot$.
The total transfer during the first 1000 slots is 1072 packets.  During this period, the source tries to send $0.48/0.08 * 1000 = 6000$ packets.
During the rest of the scenario, the system behaves, for the periods when the harvesting parameters are constant, roughly as predicted in Sec. \dref{low}.
For example, during the period from 3000 to 3500, the approximate cycle length is $I = (2 * 0.48 * 20)/((0.48)^2 - (0.40754 - 0.4)^2) = 85.442$ slots with drift
$(0.40754 + 0.4 - 0.48)/2 = 0.16337 \, mJ/slot$.  The simulation shows an 88 slot cycle with drift $13.1/88 = 0.149 \, mJ/slot$.  We recall that the analytic results are just approximations calculated under the assumptions that $c_t = c_r = 0$ and that the thresholds are exact multiples of the slot energies.

It is often the case that the node harvesting patterns are roughly known.  In this case one can design an appropriate schedule for the input.  Assuming that the destination has a transmitter that can be heard at the source, one way to implement this is for it to monitor the transfer and from time to time to transmit desirable input rates to the source.  An example of a variable input rate is $cg2/c$ of Fig. \dref{fig-Harvest}, so that the total input over the 8000 slots is the same as in the first example.  In the second scenario depicted in Fig. \dref{fig-Variable2}, the total transfer in the first 1000 slots is 977 packets, but the total input attempt is for 1000 packets only.  The final result is that the total throughput in the second scenario is 46,353 packets, whereas in the first scenario is only 41,175 packets.

\begin{figure}[hbtp]
\begin{center}
\includegraphics[scale=0.35]{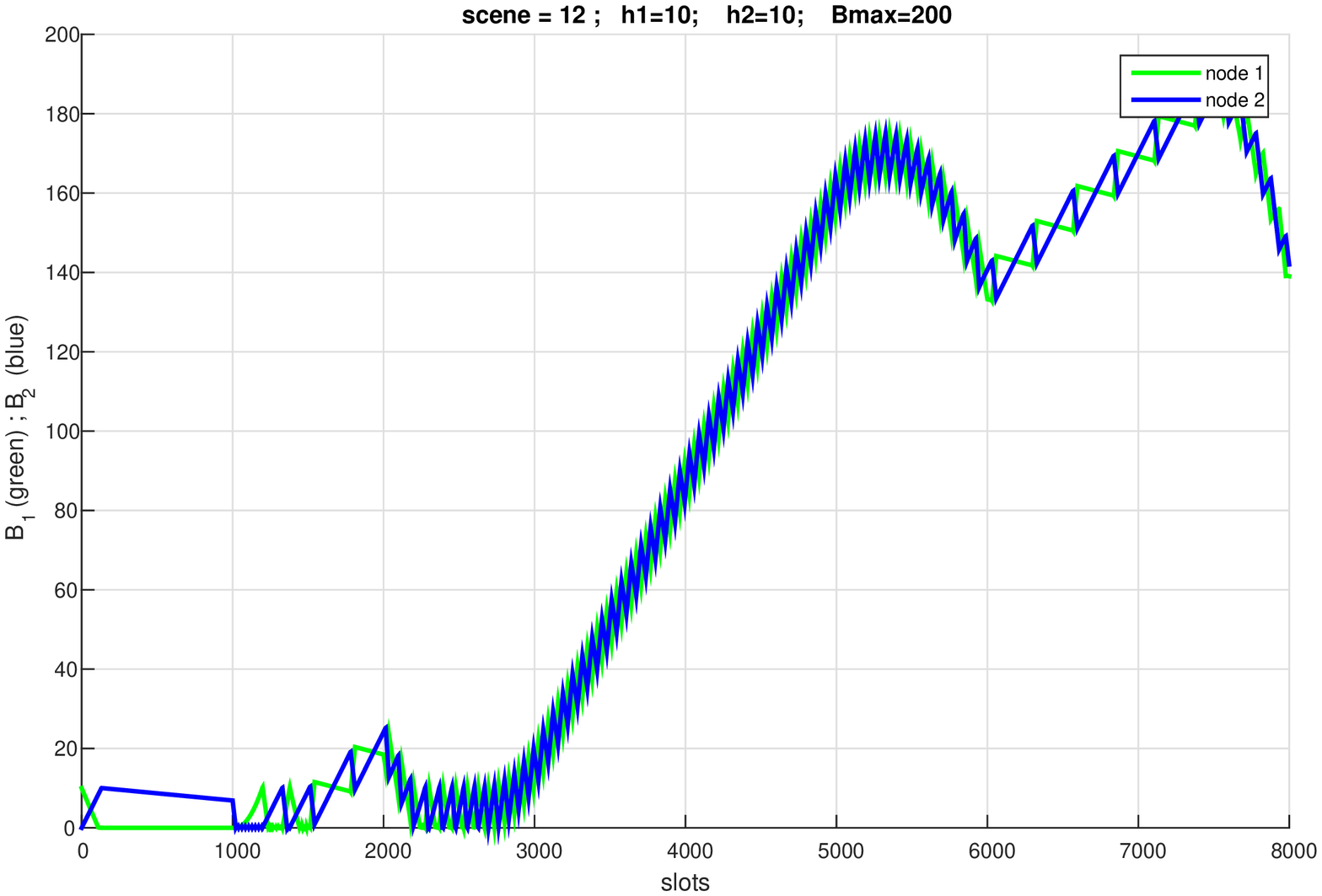}
\caption{Behavior of the System with Variable Input $cg2/c$\debug{\fbox{fig-Variable2}}\dlabel{fig-Variable2}}
\end{center}
\end{figure}

\begin{table}[hbtp]
\begin{center}
\begin{tabular}{|c|c|c|c|c|c|c|c|c|}
  \hline
input & 1000 &  5000  &  8000 & 8000 & 8000 & 8000 & 5000 &5000 \\ \hline
throughput &977 &    3830 & 7580 & 7992 & 7992 & 7992 & 4995 & 4995 \\ \hline
\end{tabular}
\caption{Inputs vs. Throughput in Scenario 2\debug{\fbox{Table-Thru2}}\dlabel{Table-Thru2}}
\end{center}
\end{table}


\newpage
\renewcommand{\fname}{} \renewcommand{\sname}{}
\bibliographystyle{alpha}

\addcontentsline{toc}{section}{References}
\begin{thebibliography}{RMME09}

\bibitem[AY]{AY} K. Akkaya and M. Younis, \emph{A survey on routing protocols for wireless sensor networks}, Ad Hoc Networks Vol.3 (2005) pp. 325–349 \debug{[survey3]}

\bibitem[ASSC]{ASSC} I.F. Akyildiz, W. Su, Y. Sankarasubramaniam, E. Cayirci, \emph{Wireless sensor networks: a survey}, Computer Networks Vol.38 (2002), pp. 393–422 \debug{[Survey1]}

\bibitem[GMSplus]{GMSplus} M.Gorlatova, R. Margolies, J. Sarik, G. Stanje, J.Zhu , B. Vigraham, M. Szczodrak, L. Carloni, P. Kinget, I. Kymissis, G. Zussman, \emph{Energy Harvesting Active Networked Tags (EnHANTs): Prototyping and Experimentation}, Columbia University, Electrical Engineering Technical Report \#2012-07-27, July 2012, available at http://enhants.ee.columbia.edu/images/papers/cu-ee-2012-07-27.pdf \debug{[Columbia Tech Report]}

\bibitem[GWZ]{GWZ}M. Gorlatova, A. Wallwater, G. Zussman,
\emph{Networking Low-Power Energy Harvesting Devices: Measurements and
  Algorithms},  Proc. IEEE INFOCOM'11, April, 2011 \debug{ [Infocom 2011]}

\bibitem[KK]{KK} J. N. Al-Karaki and A. E. Kamal, \emph{Routing Techniques in Wireless Sensor Networks: A Survey}, IEEE Wireless Communications, Dec. 2004 \debug{[survey2]}

\bibitem[LS]{LS} I. Ledvich and A. Segall, \emph{Threshold-related throughput – A new criterion for evaluation of sensor network performance}, Ad Hoc Networks, Special Issue on Recent Research Directions in Wireless Ad Hoc Networking, Vol. 5, Issue 8, November 2007, Pages 1329–1348 \debug{[Segall Ledvich]}

\bibitem[MSZ]{MSZ}  J. Marasevic, C. Stein, G. Zussman, \emph{Max-min Fair Rate Allocation and Routing in Energy Harvesting Networks: Algorithmic Analysis}, Proc. ACM MOBIHOC'14, 2014 \debug{ [MSZ- MaxMin]}














\end{thebibliography}

\end{document}